\documentclass{aa}  

\usepackage{graphicx}
\usepackage{xfrac}
\usepackage{txfonts}
\usepackage[colorlinks=true,citecolor=blue,linkcolor=blue]{hyperref}
\usepackage{amsmath}    
\usepackage{amssymb}    
\usepackage{longtable}

\newcommand{\snia}{SN~Ia}
\newcommand{\sneia}{SNe~Ia}

\let\ts=\thinspace

\newcommand{\two}{\ts {\sc ii}}
\newcommand{\three}{\ts {\sc iii}}

\newcommand{\nifs}{\ensuremath{^{56}\mathrm{Ni}}}
\newcommand{\nife}{\ensuremath{^{58}\mathrm{Ni}}}
\newcommand{\cofs}{\ensuremath{^{56}\mathrm{Co}}}
\newcommand{\feff}{\ensuremath{^{54}\mathrm{Fe}}}
\newcommand{\fefs}{\ensuremath{^{56}\mathrm{Fe}}}
\newcommand{\msun}{\ensuremath{\mathrm{M}_{\odot}}}
\newcommand{\zsun}{\ensuremath{\mathrm{Z}_{\odot}}}
\newcommand{\mtot}{\ensuremath{M_\mathrm{tot}}}

\newcommand{\kms}{\ensuremath{\mathrm{km\,s}^{-1}}}

\newcommand{\gcc}{\ensuremath{\mathrm{g\,cm}^{-3}}}

\newcommand{\mch}{\ensuremath{M_\mathrm{Ch}}}
\newcommand{\ye}{\ensuremath{Y_\mathrm{e}}}
\newcommand{\nett}{\ensuremath{^{22}\mathrm{Ne}}}

\def\cmfgen{CMFGEN}
\usepackage{color}

\makeatletter
\newcommand{\vast}{\bBigg@{4}}
\newcommand{\Vast}{\bBigg@{5}}
\makeatother

\begin{document} 

   \title{Stable nickel production in Type Ia supernovae: A smoking gun for the progenitor mass?}
   \titlerunning{[Ni\two] lines in \mch\ vs. sub-\mch\ \sneia}
   
   \author{
     S. Blondin\inst{1,2}
     \and
     E. Bravo\inst{3}
     \and
     F.~X.~Timmes\inst{4,5}
     \and
     L.~Dessart\inst{6}
     \and
     D.~J. Hillier\inst{7}
   }
   \authorrunning{S. Blondin et al.}
   
   \institute{
     Aix Marseille Univ, CNRS, CNES, LAM, Marseille, France\\
     \email{stephane.blondin@lam.fr}
     \and     
     Unidad Mixta Internacional Franco-Chilena de Astronom\'ia,
     CNRS/INSU UMI 3386 and Instituto de Astrof\'isica,\\
     Pontificia Universidad Cat\'olica de Chile, Santiago, Chile
     \and
     E.T.S. Arquitectura del Vall\`es, Universitat Polit\`ecnica de
     Catalunya, Carrer Pere Serra 1-15, 08173 Sant Cugat del Vall\`es,
     Spain
     \and
     School of Earth and Space Exploration, Arizona State University,
     Tempe, AZ, USA
     \and
     Joint Institute for Nuclear Astrophysics Center for the Evolution
     of the Elements, USA
     \and
     Institut d'Astrophysique de Paris, CNRS-Sorbonne Université, 98
     bis boulevard Arago, 75014, Paris, France
     \and
     Department of Physics and Astronomy \& Pittsburgh Particle
     Physics, Astrophysics, and Cosmology Center (PITT PACC),\\
     University of Pittsburgh, 3941 O’Hara Street, Pittsburgh, PA
     15260, USA
   }

   \date{Received 28 September 2021; accepted 26 January 2022}

  \abstract
   {At present, there are strong indications that white dwarf (WD) stars with
     masses well below the Chandrasekhar limit ($\mch\approx1.4$\,\msun)
     contribute a significant fraction of \snia\ progenitors. The
     relative fraction of stable iron-group elements 
     synthesized in the explosion has been suggested as a possible
     discriminant between \mch\ and sub-\mch\ events. In particular,
     it is thought that the higher-density ejecta of \mch\ WDs, which
     favours the synthesis of stable isotopes of nickel,
     results in prominent [Ni\two] lines in late-time spectra
     ($\gtrsim 150$\,d past explosion).}
   {We study the explosive nucleosynthesis of stable nickel in
     \sneia\ resulting from \mch\ and sub-\mch\ progenitors.
     We explore the potential for lines of [Ni\two] in the optical an
     near-infrared (at 7378\,\AA\ and 1.94\,$\mu$m) in late-time
     spectra to serve as a diagnostic of the exploding WD mass.} 
   {We reviewed stable Ni yields across a large variety of published
     \snia\ models. Using 1D \mch\ delayed-detonation and
     sub-\mch\ detonation models, we studied the synthesis of stable Ni
     isotopes (in particular, \nife) and investigated the formation of
     [Ni\two] lines using 
     non-local thermodynamic equilibrium radiative-transfer
     simulations with the \cmfgen\ code.} 
   {We confirm that stable Ni production is generally more efficient
     in \mch\ explosions at solar metallicity (typically 0.02--0.08\,\msun\ for the
     \nife\ isotope), but we note that the \nife\ yield in
     sub-\mch\ events systematically exceeds 0.01\,\msun\ for WDs that are more
     massive than one solar mass. We find that the radiative
     proton-capture reaction
     $^{57}\mathrm{Co}(\mathrm{p},\gamma)^{58}\mathrm{Ni}$ is the
     dominant production mode for \nife\ in both \mch\ and
     sub-\mch\ models, while the $\alpha$-capture reaction on \feff\ has a
     negligible impact on the final \nife\ yield.   
     More importantly, we demonstrate that the lack of [Ni\two] lines
     in late-time spectra of sub-\mch\ events is not always due to an
     under-abundance of stable Ni; rather, it results from the 
     higher ionization of Ni in the inner ejecta. Conversely, the
     strong [Ni\two] lines predicted in our 1D \mch\ models are
     completely suppressed when \nifs\ is sufficiently mixed with the
     innermost layers, which are rich in stable iron-group elements.}
   {[Ni\two] lines in late-time \snia\ spectra have a
     complex dependency on the abundance of stable Ni, which limits
     their use in distinguishing among \mch\ and
     sub-\mch\ progenitors. However, we argue that a
       low-luminosity \snia\ displaying strong [Ni\two] lines would
       most likely result from a Chandrasekhar-mass progenitor.}


   \maketitle


\section{Introduction}

In the long-standing model for Type Ia supernovae (\sneia), a
carbon-oxygen white dwarf (CO WD) star accretes material from a binary
companion until it approaches the Chandrasekhar-mass limit for a
relativistic degenerate electron plasma ($\mch \approx
1.4$\,\msun). While this model provides a robust ignition mechanism
for runaway carbon fusion, it is in tension with the observed
\snia\ rate (see e.g. \citealt{Maoz/Mannucci:2012};
\citealt{Livio/Mazzali:2018}; \citealt{WangB:2018};
\citealt{Soker:2019} for reviews). Moreover, there is growing evidence
that it cannot explain the full range of observed \snia\ properties
\citep[e.g.][]{Jha/etal:2019}. For instance, \cite{Floers/etal:2020}
find that the Ni/Fe abundance ratio inferred from late-time
spectroscopy is consistent with the predictions of sub-\mch\ models
for 85\% of normal \sneia.

There are multiple paths leading to the explosion of a WD
significantly below the Chandrasekhar-mass limit. In double-detonation
models
\citep[e.g.][]{Shen/etal:2018,Townsley/etal:2019,Magee/etal:2021,Gronow/etal:2021},
a sub-\mch\ WD accretes a thin He-rich layer from a non-degenerate
binary companion, triggering a detonation at its base which leads to a
secondary detonation of the CO core. Modern incarnations of this model
consider modestly CO-enriched low-mass He layers ($\lesssim
10^{-2}$\,\msun), whose detonation does not lead to spurious
spectroscopic features from iron-group elements (IGEs) at early times
\citep{Shen/Moore:2014,Townsley/etal:2019}. Furthermore, the predicted
rate of double-detonation models matches the observed \snia\ rate
\citep[e.g.][]{Ruiter/etal:2011,Ruiter/etal:2014}.

Other sub-\mch\ progenitor models involve double-WD systems. In the
classical double-degenerate model of \cite{Webbink:1984}, two
unequal-mass WDs in a close binary system merge through loss of energy
and angular momentum via gravitational-wave radiation. The more
massive WD tidally disrupts and accretes the lower-mass object,
resulting in an off-centre carbon ignition in the merger remnant and
the formation of an oxygen-neon (ONe) WD
\citep[e.g.][]{Saio/Nomoto:1985,Timmes/etal:1994,Shen/etal:2012}. If
the remnant mass exceeds the Chandrasekhar limit, accretion-induced
collapse to a neutron star ensues, associated with a weak explosion,
but with no \snia\ event \citep{Nomoto/Kondo:1991}.

This model was later revised by \cite{Pakmor/etal:2010}, who
considered the merger of two nearly equal-mass WDs, in which the less
massive WD is rapidly accreted onto the primary WD, resulting in
compressional heating of the accreted material and subsequent carbon
ignition. Such violent merger models have been successful in
reproducing the observed properties of both sub-luminous and normal
\sneia\ \citep{Pakmor/etal:2010,Pakmor/etal:2012}, as well as more
peculiar events \citep{Kromer/etal:2013b,Kromer/etal:2016}.

In addition to mergers, several authors have explored collisions
between two WDs as a potential \snia\ progenitor scenario
\citep[e.g.][]{Raskin/etal:2009,Rosswog/etal:2009}. Such collisions
are expected to occur in dense stellar environments such as globular
clusters \citep[e.g.][]{Hut/Inagaki:1985,Sigurdsson/Phinney:1993}; it
was more recently suggested that they may efficiently occur in triple
systems to explain \sneia\ \citep{Katz/Dong:2012,Kushnir/etal:2013},
although the predicted rates vary significantly \citep[see
  e.g.][]{Toonen/etal:2018}. Furthermore, \cite{Dong/etal:2015} argued
that the doubly peaked line profiles observed in late-time spectra of
several \sneia\ result from a bimodal \nifs\ distribution produced in
WD-WD collisions.

It should be possible to identify \sneia\ resulting from
  \mch\ vs. sub-\mch\ progenitors observationally, since 
variations in the ejecta mass have an impact on the radiative display
\citep[e.g.][]{Pinto/Eastman:2000a}. The typical photon diffusion
time depends on the mean opacity, $\kappa$, of the ejecta, its mass,
$M_\mathrm{ej}$, and characteristic velocity, $\varv$, following
$t_\mathrm{diff} \propto \kappa^{1/2} M_\mathrm{ej}^{1/2}
\varv^{-1/2}$
\citep[e.g.][]{Arnett:1982a,Woosley/etal:2007,Piro/etal:2010,Khatami/Kasen:2019}. The
photon diffusion time is thus shorter for a sub-\mch\ ejecta compared
to a \mch\ ejecta, resulting in shorter bolometric rise times
\citep[see e.g.][]{Blondin/etal:2017}. The post-maximum bolometric
decline is also faster, as the lower density of sub-\mch\ ejecta
favours the earlier escape of $\gamma$-rays
\citep[e.g.][]{Kushnir/etal:2020,Sharon/Kushnir:2020}\footnote{We
  note, however, that both of these studies argue that observed
  \snia\ light curves are in tension with the $\gamma$-ray escape time
  scales inferred for sub-\mch\ models.}.

The colour evolution around maximum light is also affected by the
ejecta mass. As noted by \cite{Blondin/etal:2017}, sub-\mch\ ejecta
are subject to a larger specific heating rate at maximum
light\footnote{defined as $\dot{e}_\mathrm{decay}(t_\mathrm{max}) =
  L_\mathrm{decay}(t_\mathrm{max})/M_\mathrm{tot}$, where
  $L_\mathrm{decay}(t_\mathrm{max})$ is the decay luminosity at
  maximum light and \mtot\ is the ejecta mass.} for a given
\nifs\ mass, owing to the shorter rise times and lower ejecta mass;
hence, they display bluer maximum-light colours. The colour evolution
past maximum light is also more pronounced, as observed in
low-luminosity
\sneia\ \citep[e.g.][]{Blondin/etal:2017,Shen/etal:2021}.  The ejecta
mass can, in principle, be constrained based purely on photometric
indicators. In practice, however, their interpretation is subject to
uncertainties in the radiative-transfer modelling \citep[see e.g.
  discussion in][]{Blondin/etal:2018}.

A more robust signature of the WD mass should therefore be sought in
the spectroscopic signatures of distinct abundance patterns predicted
by different explosion models; in particular, the density at which the
CO fuel is ignited affects the resulting nucleosynthesis. More
specifically, Chandrasekhar-mass WDs have central densities
$\rho_\mathrm{c} \gtrsim 10^9$\,\gcc, where explosion models involving
sub-\mch\ WDs (including WD mergers and collisions) detonate the CO
core in regions with $\rho \lesssim 10^8$\,\gcc. The higher densities
in \mch\ models result in a higher electron-capture rate during the
explosion, which enhances the production of neutron-rich stable
isotopes of iron-group elements compared to sub-\mch\ models.

Of particular interest are the stable isotopes of nickel, the most
abundant of which is \nife. At sufficiently late times ($\gtrsim
150$\,d past explosion), \snia\ spectra are dominated by forbidden
lines of singly and doubly ionized Ni, Co, and Fe. By then, the only
nickel left in the ejecta is stable Ni synthesized in the explosion,
whereas most of the Co is \cofs\ from \nifs\ decay, and Fe is a
mixture of primordial stable Fe and \fefs\ from \nifs\ decay. The
larger abundance of stable Ni in \mch\ models is thus expected to
manifest itself in the form of forbidden lines of [Ni\two], which have
been detected in late-time spectra of several \sneia\ to date
\citep[e.g.][]{Dhawan/etal:2018,Maguire/etal:2018, Floers/etal:2018,
  Floers/etal:2020}.  These lines ought to be largely suppressed if
not completely absent from sub-\mch\ models due to the lower abundance
of stable Ni.

In principle, we thus have a clear prediction in terms of stable Ni
production that depends on the mass of the exploding WD and an
associated spectroscopic diagnostic to distinguish between \mch\ and
sub-\mch\ models. This was partly confirmed by
\cite{Blondin/etal:2018} for low-luminosity \snia\ models: the
\mch\ model displayed a prominent line due to [Ni\two] 1.94\,$\mu$m,
where the sub-\mch\ model showed no such line.

In this paper, we test whether this prediction holds for
higher-luminosity models, which correspond to the bulk of the
\snia\ population, and to what extent the abundance of stable Ni is
the determining factor in explaining the strength of [Ni\two] lines in
late-time \snia\ spectra. We first briefly present the \snia\ models
and numerical methods in Sect.~\ref{sect:num}. We investigate the
dominant nuclear reactions responsible for stable Ni production and
present a census of \nife\ yields in \mch\ versus sub-\mch\ models in
Sect.~\ref{sect:ni}. We discuss the relative impact of Ni abundance
and ionization on nebular [Ni\two] lines in Sect.~\ref{sect:ni2}, as
well as the impact of mixing in Sect.~\ref{sect:mix}. Our conclusions
follow in Sect.~\ref{sect:ccl}.


\section{Explosion models and radiative transfer}\label{sect:num}

We base our analysis on previously published \snia\ explosion
models. One exception is the 1D \mch\ delayed-detonation model
5p0\_Z0p014, published here for the first time, whose WD progenitor
results from the evolution of a 5\,\msun\ star at solar metallicity
($Z=0.014$; \citealt{Asplund/etal:2009}). The explosive phase was
simulated using the same hydrodynamics and nucleosynthesis code as the
sub-\mch\ detonation models 1p06\_Z2p25e-2 and 0p88\_Z2p25e-2 from a
1.06\,\msun\ and 0.88\,\msun\ WD progenitor, respectively, at slightly
super-solar metallicity ($Z=0.025\approx 1.6$\,\zsun;
\citealt{Bravo/etal:2019}).  Basic model properties and various nickel
isotopic abundances are given in Table~\ref{tab:all}.

The synthetic late-time spectra ($\sim 190$\,d past explosion)
presented in Sects.~\ref{sect:ni2} and \ref{sect:mix} were computed
using the 1D, time-dependent, non-local thermodynamic equilibrium
radiative-transfer code \cmfgen\ of
\cite{Hillier/Dessart:2012}. Late-time spectra for the low-luminosity
\mch\ delayed-detonation model DDC25 and the sub-\mch\ detonation
model SCH2p0 are from \cite{Blondin/etal:2018}. Those for the
high-luminosity \mch\ delayed-detonation model DDC0 and the
sub-\mch\ detonation model SCH7p0 are published here for the first
time. The same is true of the mixed versions of the
\mch\ delayed-detonation model DDC15 (Sect.~\ref{sect:mix}). All model
outputs are publicly available
online\footnote{\url{https://zenodo.org/record/5528088}}.


\section{Stable Ni production in \mch\ vs. sub-\mch\ models}\label{sect:ni}

\subsection{Nuclear statistical equilibrium}\label{sect:nse}

\begin{figure}
\centering
\includegraphics{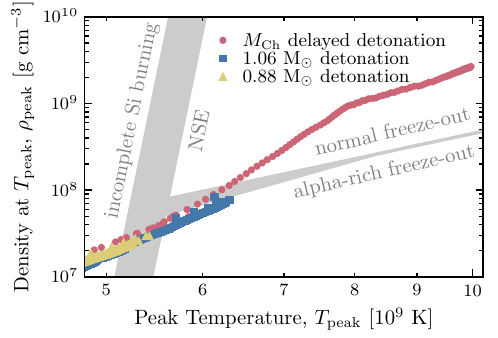}
\caption{\label{fig:trho}
Density at peak temperature ($\rho_\mathrm{peak}$) versus peak
temperature ($T_\mathrm{peak}$, in units of $10^9$\,K) in the
\mch\ delayed-detonation model 5p0\_Z0p014 (filled circles) and the
sub-\mch\ detonation models 1p06\_Z2p25e-2
($M_\mathrm{WD}=1.06$\,\msun, filled squares) and 0p88\_Z2p25e-2
($M_\mathrm{WD}=0.88$\,\msun, filled triangles). The wide vertical
band denotes the transition between incomplete Si burning and complete
burning to NSE. The NSE region is further subdivided into `normal' and
`alpha-rich' freeze-out regimes. The width of the bands correspond to
variations in the post-burn cooling time scale
(\citealt{Woosley/etal:1973}; see also \citealt{Lach/etal:2020}, their
Fig.~1).
}
\end{figure}

Explosive burning at sufficiently high temperatures ($T \gtrsim
5\times 10^9$\,K) results in a state of balance between forward and
reverse nuclear reactions known as nuclear statistical equilibrium
(NSE; see e.g.
\citealt{Clifford/Tayler:1965,Woosley/etal:1973,Hartmann/Woosley/ElEid:1985,Cabezon/Garcia-Senz/Bravo:2004,Nadyozhin/Yudin:2004}). Such
temperatures are reached in the inner layers of both
\mch\ delayed-detonation models and sub-\mch\ detonation models,
although the highest temperatures $\lesssim 10^{10}$\,K are only
reached in \mch\ models (Fig.~\ref{fig:trho}). In NSE, the yields do
not depend on the initial composition but are instead uniquely
determined by the peak temperature ($T_\mathrm{peak}$), the density at
$T_\mathrm{peak}$ ($\rho_\mathrm{peak}$), and the electron fraction,
$\ye = \sum_i (Z_i/A_i) X_i$, where $X_i$ is the mass fraction of a
particular isotope $i$ with atomic number $Z_i$ and mass number
$A_i$.

The electron fraction of the WD prior to explosion is set by the
metallicity of the progenitor star on the main sequence, which is
routinely parametrized by adjusting the abundance of the neutron-rich
isotope \nett. Here, the assumption is that the CNO catalysts
all end up as $^{14}$N at the end of the hydrogen-burning phase,
which is then converted to \nett\ via
$^{14}$N($\alpha$,$\gamma$)$^{18}$F($\beta+$,$\nu_\mathrm{e}$)$^{18}$O($\alpha$,$\gamma$)$^{22}$Ne
during the helium-core burning phase \citep[see][]{Timmes/etal:2003},
such that:

\begin{equation}
\label{eq:xne22}
X(\nett) = 22 \left[
  \frac{X(^{12}\mathrm{C})}{12} +
  \frac{X(^{14}\mathrm{N})}{14} +
  \frac{X(^{16}\mathrm{O})}{16}
  \right]
\approx 0.013 \left(\frac{Z}{\mathrm{Z}_\odot}\right),
\end{equation}

\noindent
where we used the solar CNO abundances from
\cite{Asplund/etal:2009} and isotopic ratios from
\cite{Lodders:2003}\footnote{\cite{Kushnir/etal:2020} adopt the
slightly higher value of $X(\nett)\approx 0.015(Z/\mathrm{Z}_\odot)$
to account for the expected higher solar bulk abundances compared to
the photospheric values \citep[see e.g.][]{Turcotte/Wimmer:2002}.}. We
ignore the initial \nett\ of the progenitor star as its mass fraction
is $\sim 10^{-4}$ at solar metallicity.

In addition to \nett\ resulting from the CNO cycle, the initial
metallicity is also determined by the abundance of \fefs\ nuclei
inherited from the ambient interstellar medium.  For a WD composed of
only $^{12}$C, $^{16}$O, $^{22}$Ne, and
\fefs\ (i.e. $X(^{12}\mathrm{C}) + X(^{16}\mathrm{O}) = 1 -
X(^{22}\mathrm{Ne}) - X(^{56}\mathrm{Fe})$), the electron fraction is \citep[see also][]{Kushnir/etal:2020}:

\begin{equation}
\label{eq:ne22}
\begin{split}
  \ye & = \frac{6}{12} X(^{12}\mathrm{C}) +
          \frac{8}{16} X(^{16}\mathrm{O}) +
          \frac{10}{22} X(^{22}\mathrm{Ne}) +
          \frac{26}{56} X(^{56}\mathrm{Fe}) \\
      & = \frac{1}{2} -
          \frac{X(^{22}\mathrm{Ne})}{22} -
          \frac{X(^{56}\mathrm{Fe})}{28}\\
      & \approx \frac{1}{2} - 6.5 \times 10^{-4} \left(\frac{Z}{\mathrm{Z}_\odot}\right),
\end{split}
\end{equation}

\noindent
where we have used the solar Fe abundance $X(\mathrm{Fe})=1.292\times
10^{-3}$ from \cite{Asplund/etal:2009} and the \fefs\ isotopic
fraction of 91.754\% from \cite{Lodders:2003}, yielding
$X(\fefs)\approx 1.185\times 10^{-3} (Z/\mathrm{Z}_\odot)$. In this
framework, a solar-metallicity WD has $X(^{22}\mathrm{Ne}) \approx
0.013$\footnote{\cite{Kobayashi/etal:2020} adopt a different approach
in their solar-scaled initial composition models by assuming that
all the \nett\ is inherited from the progenitor with no contribution
from CNO. This results in a much lower \nett\ mass fraction at a
given metallicity (although we were not able to confirm the exact
value), which has a significant impact on the stable Ni yields in
their models (see Sect.~\ref{sect:ni58yield}).} and $\ye\approx
0.49935$. A larger metallicity corresponds to a larger
\nett\ abundance and in turn a lower \ye.

\begin{figure}[h!]
\centering
\includegraphics{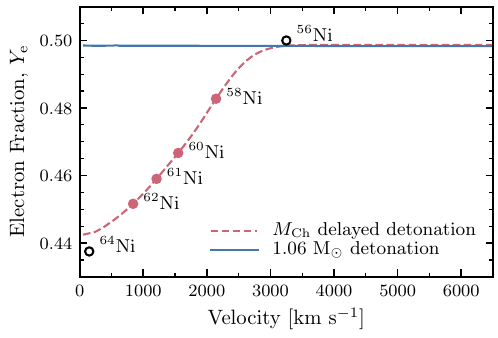}\vspace{.1cm}
\includegraphics{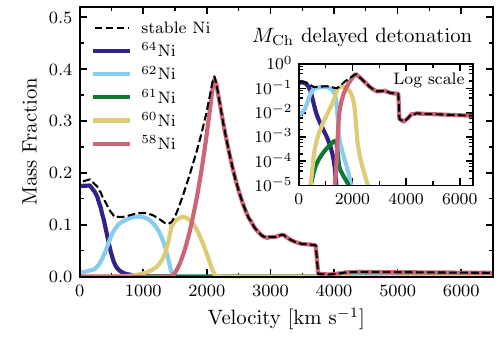}\vspace{.1cm}
\includegraphics{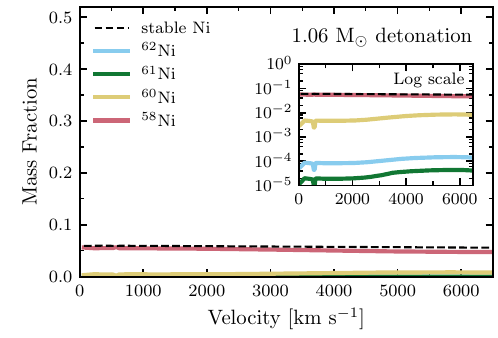}
\caption{\label{fig:yeni58}
Top panel: Electron fraction profile at $t\approx 30$\,min past
explosion in the inner ejecta of the \mch\ delayed-detonation model
5p0\_Z0p014 (dashed line) and the
$M_\mathrm{WD}=1.06$\,\msun\ sub-\mch\ detonation model 1p06\_Z2p25e-2
(solid line) shown in Fig.~\ref{fig:trho}. The markers in the upper
panel are shown such that the ordinate corresponds to the \ye\ of the
nucleus (e.g. $\ye = 28/58 \approx 0.483$ for \nife) and the abscissa
to the interpolated velocity on the \ye\ profile for the
\mch\ delayed-detonation model (both \nifs\ and $^{64}$Ni are
synthesized in this model but the \ye\ profile does not intersect the
\ye\ value of either isotope).  Middle and bottom panels: Abundance
profiles of stable Ni isotopes for both models. The insets correspond
to a logarithmic scale, revealing minor contributions to the total
stable Ni abundance from $^{61}$Ni for the \mch\ model and from both
$^{61}$Ni and $^{62}$Ni for the sub-\mch\ model. No $^{64}$Ni is
produced in this sub-\mch\ model.
}
\end{figure}

In \mch\ models this baseline \ye\ can, in principle, be reduced via
weak reactions on carbon during the convective burning (or
`simmering') phase prior to thermonuclear runaway
\citep[e.g.][]{Piro/Bildsten:2008,Chamulak/etal:2008,Schwab/etal:2017},
although \cite{MartinezRodriguez/etal:2016} show the impact to be
negligible (reduction in \ye\ of $\lesssim 10^{-4}$ ;
but see \citealt{Piersanti/etal:2017} for a different
  view).  However, the higher densities of \mch\ WDs (up to
$2\mathrm{-}3 \times 10^9$\,\gcc; see Fig.~\ref{fig:trho}) result in
a significant electron-capture rate during the initial deflagration
phase of delayed-detonation models. This lowers the \ye\ far below the
baseline value (Fig.~\ref{fig:yeni58}, top panel) and favours the
synthesis of neutron-rich isotopes in the innermost layers ($\varv
\lesssim 3000$\,\kms; Fig.~\ref{fig:yeni58}, middle
  panel). For nickel, this results in the synthesis of the stable
isotopes $^{58}$Ni, $^{60}$Ni, $^{61}$Ni, $^{62}$Ni, and $^{64}$Ni
instead of the radioactive \nifs\ (for which $\ye=0.5$).

\begin{figure}
\centering
\includegraphics{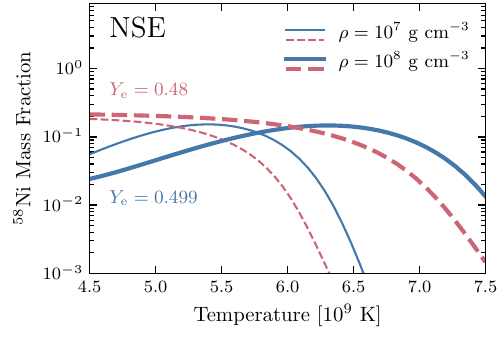}\vspace*{.25cm}
\includegraphics{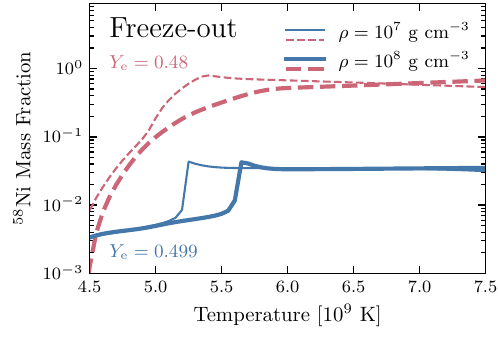}
\caption{\label{fig:ninse}
Top panel: NSE distributions for \nife\ as a function of
temperature (in units of $10^9$\,K) for $\ye = 0.48$ (dashed lines)
and $\ye = 0.499$ (solid lines), at densities of $10^7$\,\gcc\ (thin
lines) and $10^8$\,\gcc\ (thick lines).
These NSE distributions were computed with the
\texttt{public\_nse} code
\protect\footnotemark\
(see also \citealt{Seitenzahl/etal:2008}).
Bottom panel: Freeze-out yields based on the adiabatic thermodynamic
trajectories of \cite{Magkotsios/etal:2010} for the same set of
$(\rho_\mathrm{peak}, \ye)$ values.
}
\end{figure}

In detonations of sub-\mch\ WDs, however, the burning timescale is
much shorter than the weak-reaction timescale, such that \ye\ remains
constant at its baseline value ($\ye \approx 0.49935$ for a
solar-metallicity WD) throughout the burning phase. Stable
neutron-rich isotopes of nickel are still synthesized in NSE at this
\ye\ at the peak temperatures ($5\mathrm{-}6 \times 10^9$\,K) and
densities ($10^7\mathrm{-}10^8$\,\gcc), characteristic of the inner
ejecta of sub-\mch\ detonations (Fig.~\ref{fig:yeni58}, bottom panel).
These conditions are similar to those encountered in the layers of
\mch\ models where most of the \nife\ is synthesized ($\varv \approx
2000$\,\kms\ for the \mch\ model shown in Fig.~\ref{fig:yeni58}). By
comparing the NSE distributions for $\ye=0.499$ and $\ye=0.48$ in
Fig.~\ref{fig:ninse} (top panel), we see that the \nife\ abundance in
sub-\mch\ detonations can be comparable to (and even exceed) that of
\mch\ delayed-detonation models, with predicted mass fractions
$X(\nife) \approx 0.1$.

\footnotetext{Publicly available on F.~X. Timmes' webpage;
\url{http://cococubed.asu.edu/code\_pages/nse.shtml}}

\subsection{Freeze-out yields}

The final isotopic abundances can differ significantly from their NSE
value during the so-called freeze-out phase, when free
particles (protons, neutrons, $\alpha$ particles) reassemble into
nuclei on a timescale of $\propto 1 / \sqrt{\rho}$
\citep[e.g.][]{Magkotsios/etal:2010}. For high densities ($\gtrsim
10^8\mathrm{-}10^9$\,\gcc), this timescale is short and for $T \lesssim
7 \times 10^9$\,K the $\alpha$ abundance is low (normal
freeze-out), such that the final abundances do not differ greatly from
their NSE value. However, for the lower $\rho_\mathrm{peak}$ relevant to the
synthesis of \nife\ ($10^7\mathrm{-}10^8$\,\gcc),  the
freeze-out timescale is longer and the $\alpha$ abundance is higher
($X(^{4}\mathrm{He}) \approx 0.1$; alpha-rich freeze-out), and the
final yields of stable Ni isotopes can differ significantly from their
NSE values (Fig.~\ref{fig:ninse}, bottom panel).

In either case, the freeze-out timescale is significantly shorter than
the weak-reaction timescale, such that \ye\ remains roughly constant
during the freeze-out. Stable isotopes of Ni are synthesized
preferentially in shells with a similar \ye\ value. This is
illustrated by the \mch\ delayed-detonation model in the bottom panel
of Fig.~\ref{fig:yeni58}, where the peak in the stable Ni abundance
profile at $\sim 2000$\,\kms\ consists almost exclusively of the
\nife\ isotope (see inset). The $\ye$ of this isotope ($\ye = 28/58
\approx 0.48$) coincides with the $\ye$ value at this velocity
coordinate (Fig.~\ref{fig:yeni58}, top panel), where the
peak temperature
is $\sim 5 \times 10^9$\,K and the density $2\mathrm{-}3\times
10^7$\,\gcc. Under these conditions, the predicted freeze-out
mass fraction for \nife\ is a few times 0.1
(Fig.~\ref{fig:ninse}, bottom panel), which is on par with the \mch\ model
yield.  The more neutron-rich isotopes $^{60}$Ni, $^{61}$Ni, $^{62}$Ni,
and $^{64}$Ni are synthesized at lower velocities $\lesssim
1500$\,\kms,\ where the density is higher -- and hence the \ye\ value
is lower as a result of electron captures.

For the sub-\mch\ detonation model, the stable Ni mass fraction remain
roughly constant at $5\mathrm{-}6 \times 10^{-2}$ throughout the inner
ejecta. These mass fractions are in agreement with the predicted
freeze-out \nife\ yields for $\ye = 0.499$ and densities in the range
$10^7\mathrm{-}10^8$\,\gcc\ for peak temperatures $5\mathrm{-}6 \times
10^9$\,K (Fig.~\ref{fig:ninse}, bottom panel).  In \mch\ models,
\nife\ is synthesized in layers with similar
$(T_\mathrm{peak},\rho_\mathrm{peak})$ conditions, but at a lower
$\ye\approx 0.48$, which results in an order-of-magnitude difference
in the predicted freeze-out yields.

\subsection{Decayed yields at one year past explosion}

Once the burning phase has ceased (typically a few seconds after the
beginning of the thermonuclear runaway), the stable Ni yield continues
to increase through radioactive decays with half-lives
$t_{\sfrac{1}{2}}\gtrsim 1$\,s, in particular via the $\beta+$ decay
chains $^{60,61,62}$Zn $\rightarrow ^{60,61,62}$Cu
$\rightarrow^{60,61,62}$Ni (see Table~\ref{tab:decay2ni}).  From this point, we go on to consider the decayed yields at one year past explosion when
referring to the \nife\ or stable Ni yields, as these are most
relevant to the late-time spectra discussed in this paper. We note,
however, that the decayed \nife\ yield is set shortly after explosion,
since the only parent isotope, $^{58}$Cu, decays to \nife\ with a half-life
of $\sim$3.2\,s.

\begin{table}
\footnotesize
\caption{Radioactive decay chains with half-lives $t_{\sfrac{1}{2}} >
  1$\,s ending in a stable isotope of nickel.}\label{tab:decay2ni}
\begin{tabular}{c@{\hspace{2.6mm}}c@{\hspace{2.6mm}}c@{\hspace{2.6mm}}c@{\hspace{2.6mm}}c@{\hspace{2.6mm}}c}
\hline
\multicolumn{1}{c}{Parent} & \multicolumn{1}{c}{Daughter} & \multicolumn{1}{c}{Half-life}       & Decay Mode     & Final   \\
\multicolumn{1}{c}{Isotope}& \multicolumn{1}{c}{Isotope}  & \multicolumn{1}{c}{$t_{\sfrac{1}{2}}$} & [Branching \%] & Product \\
\hline
$^{58}$Cu & $^{58}$Ni & 3.204(7)\,s              & $\epsilon + \beta+$ & $^{58}$Ni \\
$^{60}$Mn & $^{60}$Fe & 51(6)\,s                 & $\beta-$            & $^{60}$Ni \\
$^{60}$Fe & $^{60}$Co & $2.62(4) \times 10^6$\,y & $\beta-$            & $^{60}$Ni \\
$^{60}$Co & $^{60}$Ni & 5.2714(6)\,y             & $\beta-$            & $^{60}$Ni \\
$^{60}$Zn & $^{60}$Cu & 2.38(5)\,m               & $\epsilon + \beta+$ & $^{60}$Ni \\
$^{61}$Fe & $^{61}$Co & 5.98(6)\,m               & $\beta-$            & $^{61}$Ni \\
$^{61}$Co & $^{61}$Ni & 1.650(5)\,h              & $\beta-$            & $^{61}$Ni \\
$^{61}$Zn & $^{61}$Cu & 89.1(2)\,s               & $\epsilon + \beta+$ & $^{61}$Ni \\
$^{61}$Cu & $^{61}$Ni & 3.333(5)\,h              & $\epsilon + \beta+$ & $^{61}$Ni \\
$^{62}$Fe & $^{62}$Co & 68(2)\,s                 & $\beta-$            & $^{62}$Ni \\
$^{62}$Co & $^{62}$Ni & 1.50(4)\,m               & $\beta-$            & $^{62}$Ni \\
$^{62}$Zn & $^{62}$Cu & 9.186(13)\,h             & $\epsilon + \beta+$ & $^{62}$Ni \\
$^{62}$Cu & $^{62}$Ni & 9.74(2)\,m               & $\epsilon + \beta+$ & $^{62}$Ni \\
$^{64}$Fe & $^{64}$Co & 2.0(2)\,s                & $\beta-$            & $^{64}$Ni \\
$^{64}$Co & $^{64}$Ni & 0.30(3)\,s               & $\beta-$            & $^{64}$Ni \\
$^{64}$Cu & $^{64}$Ni & 12.700(2)\,h             & $\epsilon + \beta+$ [61.0(3)] &$^{64}$Ni \\
\hline
\end{tabular}
\flushleft
\begin{footnotesize}
  \textbf{Notes:}
  Data are from \cite{toi}
  except for the half-life for $^{60}$Fe$\rightarrow ^{60}$Co, which
  is taken from \cite{Rugel/etal:2009}.
  Numbers in parentheses are the $1\sigma$ uncertainty on the last
  digit.  Numbers in square brackets give the branching ratio (\%),
  which is implicitly 100\% when not given.  The decay mode $\epsilon$
  refers to electron capture (EC).  Although its half-life is less
  than 1\,s, we report the $^{64}$Co$\rightarrow ^{64}$Ni decay since
  it is part of the $^{64}$Fe$\rightarrow ^{64}$Co$\rightarrow
  ^{64}$Ni decay chain.  We do not consider excited nuclear isomer
  states of $^{60\mathrm{m}}$Mn, $^{60\mathrm{m}}$Co, or
  $^{62\mathrm{m}}$Co.
\end{footnotesize}
\end{table}

\subsection{Synthesis of the stable isotope \nife}

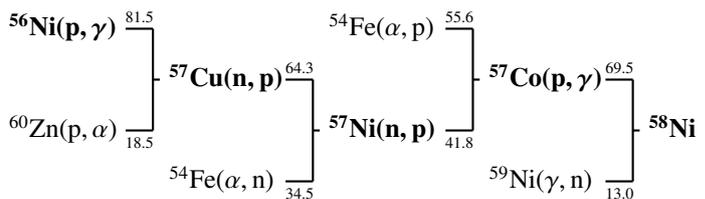
\begin{figure}
\centering

\setlength{\unitlength}{.01\linewidth}
\begin{picture}(100,35)
  \thicklines
  
  \put( 0.0, 27.0){$\boldsymbol{\nifs(\mathrm{p},\gamma)}$}
  \put( 0.0, 12.0){$^{60}\mathrm{Zn}(\mathrm{p},\alpha)$}
  \put(17.0, 28.0){\line(1,0){4}}
  \put(17.0, 13.0){\line(1,0){4}}
  \put(17.0, 29.0){{\fontsize{6}{7.2}\selectfont 81.5}}
  \put(17.0, 10.5){{\fontsize{6}{7.2}\selectfont 18.5}}
  \put(21.0, 13.0){\line(0,1){15}}
  \put(21.0, 20.5){\line(1,0){1}}

  \put(23.4, 19.5){$\boldsymbol{^{57}\mathrm{Cu}(\mathrm{n},\mathrm{p})}$}
  \put(23.4,  4.5){$^{54}\mathrm{Fe}(\alpha,\mathrm{n})$}
  \put(40.4, 20.5){\line(1,0){4}}
  \put(40.4,  5.5){\line(1,0){4}}
  \put(40.4, 21.5){{\fontsize{6}{7.2}\selectfont 64.3}}
  \put(40.4,  3.0){{\fontsize{6}{7.2}\selectfont 34.5}}
  \put(44.4,  5.5){\line(0,1){15}}
  \put(44.4, 13.0){\line(1,0){1}}

  \put(46.8, 12.0){$\boldsymbol{^{57}\mathrm{Ni}(\mathrm{n},\mathrm{p})}$}
  \put(46.8, 27.0){$^{54}\mathrm{Fe}(\alpha,\mathrm{p})$}
  \put(63.8, 13.0){\line(1,0){4}}
  \put(63.8, 28.0){\line(1,0){4}}
  \put(63.8, 10.5){{\fontsize{6}{7.2}\selectfont 41.8}}
  \put(63.8, 29.0){{\fontsize{6}{7.2}\selectfont 55.6}}
  \put(67.8, 13.0){\line(0,1){15}}
  \put(67.8, 20.5){\line(1,0){1}}

  \put(70.2, 19.5){$\boldsymbol{^{57}\mathrm{Co}(\mathrm{p},\gamma)}$}
  \put(70.2,  4.5){$^{59}\mathrm{Ni}(\gamma,\mathrm{n})$}
  \put(87.2, 20.5){\line(1,0){4}}
  \put(87.2,  5.5){\line(1,0){4}}
  \put(87.2, 21.5){{\fontsize{6}{7.2}\selectfont 69.5}}
  \put(87.2,  3.0){{\fontsize{6}{7.2}\selectfont 13.0}}
  \put(91.2,  5.5){\line(0,1){15}}
  \put(91.2, 13.0){\line(1,0){1}}

  \put(93.6, 12.0){$\boldsymbol{\nife}$}
  
\end{picture}

\caption{\label{fig:ni58chain}
Main reactions resulting in the synthesis of stable \nife. The numbers
give the probability for a given reaction product to result from a
specific reactant in the sub-\mch\ model 1p06\_Z2p25e-2 (see
  text for details).
}
\end{figure}

For all the models studied here with a \nife\ yield larger than
0.01\,\msun, more than 60\% of all the stable Ni is in the form of
\nife\ (see Table~\ref{tab:all})\footnote{The only exception is the
  double-detonation model of \cite{Townsley/etal:2019}, with a
  \nife\ isotopic fraction of $\sim 44.7$\%. A large fraction of the
  stable nickel in this model is in the form of $^{60}$Ni (44.9\%)
  which results from the radioactive decay of $^{60}$Zn.}. This
fraction rises above 80\% for models with a \nife\ yield larger than
0.04\,\msun. For comparison, the isotopic
fraction of \nife\ on Earth and in the Sun is $\sim 68$\%
\citep{Lodders:2003}.  In both the \mch\ delayed-detonation model
5p0\_Z0p014 and the sub-\mch\ detonation model 1p06\_Z2p25e-2 (described
in the previous section), this isotope is mainly synthesized through
the reactions depicted in Fig.~\ref{fig:ni58chain}, where the reaction
probabilities correspond to the aforementioned sub-\mch\ model. They
were obtained by integrating the net reaction fluxes (mol g$^{-1}$)
and calculating the relative contributions of each reaction to the
total net flux. Thus, in this model,
$^{57}$Cu is synthesized 81.5\% of the time via
$\nifs(\mathrm{p},\gamma)$ and the remaining 18.5\% via
$^{60}\mathrm{Zn}(\mathrm{p},\alpha)$. The
probabilities do not always add up to 100\%, as other
more minor reactions can contribute to a specific isotope. Most
notably \nife\ is also synthesized via 
$^{58}\mathrm{Co}(\mathrm{p},\mathrm{n})$ (6.13\%),
$^{61}\mathrm{Cu}(\mathrm{p},\alpha)$ (4.85\%),
$^{55}\mathrm{Fe}(\alpha,\mathrm{n})$ (2.26\%),
$^{54}\mathrm{Fe}(\alpha,\gamma)$ (2.04\%),
$^{57}\mathrm{Ni}(\mathrm{n},\gamma)$ (1.25\%),
$^{58}\mathrm{Cu}(\mathrm{n},\mathrm{p})$ (0.51\%),
$^{62}\mathrm{Zn}(\gamma,\alpha)$ (0.38\%), and
$^{61}\mathrm{Zn}(\mathrm{n},\alpha)$ (0.01\%), in addition to the two
dominant reactions $^{57}\mathrm{Co}(\mathrm{p},\gamma)$ (69.5\%) and
$^{59}\mathrm{Ni}(\gamma,\mathrm{n})$ (13.0\%).

The dominant reaction chain is highlighted in bold:
\nifs(p,$\gamma$)$^{57}$Cu(n,p)$^{57}$Ni(n,p)$^{57}$Co(p,$\gamma$)\nife.
The probability for a \nife\ nucleus to result from a particular
reactant nucleus is simply obtained by multiplying the reaction
probabilities along the chain. Thus $0.815 \times 0.643 \times 0.418
\times 0.695 \approx 0.152$ or 15.2\%\ of \nife\ nuclei result from
the reaction chain starting with \nifs\ in this model. The majority of
\nifs\ does not end up as \nife,\ of course, as many \nifs\ nuclei
survive the explosive phase to later decay radioactively via the
$\nifs\rightarrow\mathrm{\cofs}\rightarrow\mathrm{\fefs}$ chain and
power the \snia\ light curve.

Among all reactions ending in \nife, the final \nife\ abundance is
mostly determined by the rate for the radiative proton-capture
reaction $^{57}\mathrm{Co}(\mathrm{p},\gamma)^{58}\mathrm{Ni}$. This
might appear surprising since an $\alpha$-rich freeze-out from NSE
should favour $\alpha$ captures on \feff\ as the main production route
for \nife, in particular
$^{54}\mathrm{Fe}(\alpha,\gamma)^{58}\mathrm{Ni}$. However, this
reaction only contributes $\sim 2$\% of the net nucleosynthetic flux
to \nife\ in this model.  In spite of the seemingly large contribution
of $^{54}\mathrm{Fe}(\alpha,\mathrm{p})$ to the synthesis of $^{57}$Co
(34.5\%) and of $^{54}\mathrm{Fe}(\alpha,\mathrm{n})$ to that of
$^{57}$Ni (55.6\%), artificially inhibiting all $\alpha$ captures
(with either $\gamma$, neutron or proton output channels) on $^{54}$Fe
has a negligible impact on the final \nife\ abundance compared to when
the radiative proton-capture reaction
$^{57}\mathrm{Co}(\mathrm{p},\gamma)^{58}\mathrm{Ni}$ is artificially
switched off.

Nonetheless, the preferred reaction chain from $^{54}$Fe to
\nife\ during freeze-out mimics the $\alpha$-capture reaction
$^{54}\mathrm{Fe}(\alpha,\gamma)^{58}\mathrm{Ni,}$ as it proceeds first
via two radiative proton captures to \nifs,
namely, $^{54}\mathrm{Fe}$(p,$\gamma$)$^{55}\mathrm{Co}$(p,$\gamma$)\nifs,
followed by the reaction chain outlined in bold above from \nifs\ to
\nife. The entire process then consists of four (p,$\gamma$) and two
(n,p) reactions, which is indeed equivalent to the net capture of an
$\alpha$ particle (2n + 2p). While the abundance of $\alpha$ particles
is large in $\alpha$-rich freeze-out by definition, the abundance of
free neutrons and protons is even more enhanced compared to that in a
normal freeze-out.

Although the $^{57}\mathrm{Co}(\mathrm{p},\gamma)^{58}\mathrm{Ni}$
reaction dominates the reaction chain ending in \nife, the synthesis
of this isotope in Chandrasekhar-mass WDs is mostly affected by the
amount of electron captures in NSE. Since \nifs\ is the most abundant
isotope when NSE is achieved at the neutron excess inherited from the
progenitor (see Fig.~\ref{fig:yeni58}), the yield of \nife\ is most
sensitive to the electron-capture rate on \nifs\ and, to a lesser
extent, on $^{55}$Co. However, the yield of \nife\ is quite robust as
it changes by $\sim$20\% for a two orders-of-magnitude change in any
electron-capture rate (see \citealt{Bravo:2019} for more details).

\subsection{\nife\ yields in \mch\ and sub-\mch\ models}\label{sect:ni58yield}

\begin{figure*}
\centering
\includegraphics{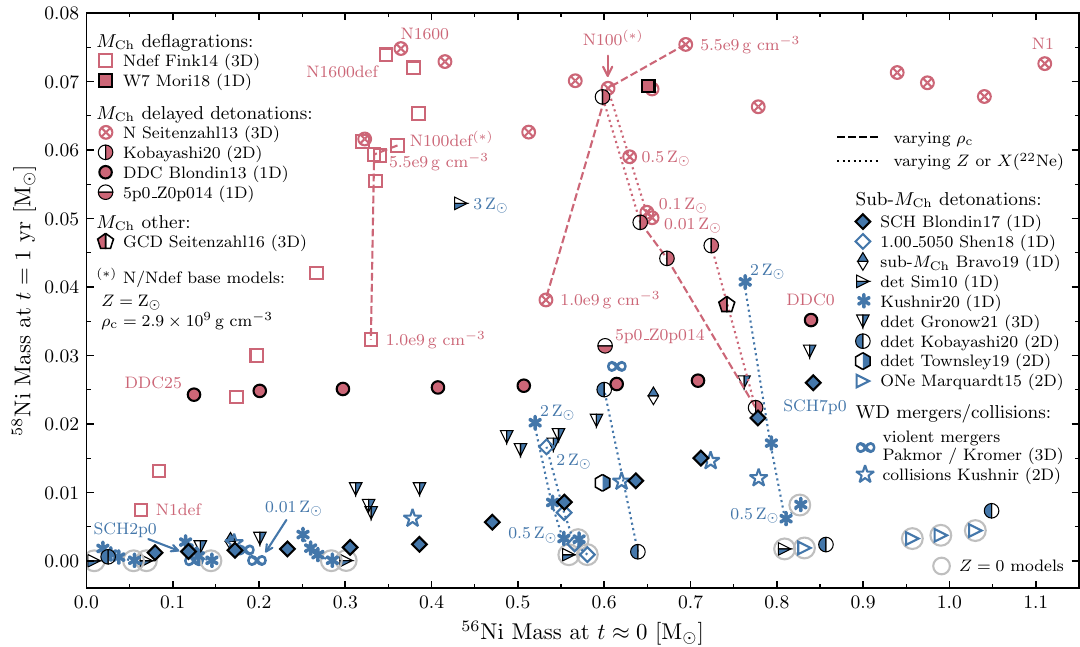}
\caption{\label{fig:mni58} Stable \nife\ yield at $t=1$\,yr past
  explosion versus radioactive \nifs\ yield at $t\approx0$ for various
  \snia\ explosion models (\mch\ models in red, sub-\mch\ models in
  blue). Chandrasekhar-mass models include: deflagrations (3D models
  of \citealt{Fink/etal:2014}; 1D W7 model of
  \citealt{Mori/etal:2018}), delayed detonations (3D models of
  \citealt{Seitenzahl/etal:2013}; 2D models of
  \citealt{Kobayashi/etal:2020}; 1D models of
  \citealt{Blondin/etal:2013}; 1D model 5p0\_Z0p014 from this paper),
  and gravitationally confined detonations (3D model of
  \citealt{Seitenzahl/etal:2016}).  Sub-\mch\ models include:
  detonations (1D models of \citealt{Blondin/etal:2017}; 1D
  1\,\msun\ models of \citealt{Shen/etal:2018}; 1D models of
  \citealt{Bravo/etal:2019}; 1D models of \citealt{Sim/etal:2010}; 1D
  models of \citealt{Kushnir/etal:2020}), double detonations (3D
  models of \citealt{Gronow/etal:2021}; 2D models of
  \citealt{Kobayashi/etal:2020}; 2D model of
  \citealt{Townsley/etal:2019}), detonations in ONe WDs (2D models of
  \citealt{Marquardt/etal:2015}), violent WD mergers (3D models of
  \citealt{Pakmor/etal:2011,Pakmor/etal:2012} and
  \citealt{Kromer/etal:2013b,Kromer/etal:2016}), and WD-WD collisions
  (2D models of Kushnir 2021, private communication).}
\end{figure*}

We show the decayed \nife\ yield at $t=1$\,yr past explosion as a
function of the \nifs\ mass at $t\approx 0$ for a variety of
\snia\ explosion models in Fig.~\ref{fig:mni58}. In the following
subsections, we first discuss \mch\ models and then
sub-\mch\ models. We include violent WD mergers and WD collisions in
sub-\mch\ models since the mass of the exploding WD is below \mch,
despite the combined mass of both WDs sometimes exceeding this value.

\subsubsection{Chandrasekhar-mass models}

\paragraph{\textbf{Deflagrations:}}

Laminar flames in \sneia\ quickly become turbulent as buoyant hot
ashes rise through overlying cold fuel, generating Rayleigh-Taylor and
Kelvin-Helmholtz instabilities that increase the flame surface and
hence the rate of fuel consumption. Deflagrations are thus best
studied in 3D, and we base our discussion on the models of
\cite{Fink/etal:2014}\footnote{For models that fail to completely
unbind the WD, the reported yields also include the \nife\ and
\nifs\ synthesized in the remnant core.}. Since the precise initial
conditions at the onset of thermonuclear runaway remain unknown to a
large extent, the deflagration is artificially ignited in a number
$N_k$ of spherical ignition spots (or `kernels') simultaneously. In
their study, \cite{Fink/etal:2014} consider $N_k = 1, 3, 5, 10, 20, 40,
100, 150, 200, 300, \ \mathrm{and}\ 1600$ such kernels distributed at
random around the WD centre.

As $N_k$ increases, the rate of fuel consumption (and hence nuclear
energy release) also increases, resulting in a more rapid flame growth
and more material being burnt. Thus, increasing $N_k$ results in a
higher production of both stable \nife\ and radioactive \nifs, which
explains the monotonic sequence in Fig.~\ref{fig:mni58} (open squares)
up until $N_k = 150$. For higher $N_k$, the nuclear energy release is
so high early on that the resulting WD expansion quenches nuclear
reactions, resulting in less material being burnt. Thus, while the
\nife\ yield continues to increase with $N_k$ (as this stable isotope
is mainly produced during the initial stages of the deflagration), the
\nifs\ yield remains more or less constant, even decreasing slightly
for $N_k = 1600$ (model N1600def).

The models of \cite{Fink/etal:2014} enable us to study the impact of
variations in the central density of the progenitor WD for model
N100def (dashed line in Fig.~\ref{fig:mni58}). A lower central density of
$\rho_\mathrm{c} = 1.0 \times 10^9$\,\gcc\ results in a $\sim 50$\%\ lower
\nife\ yield owing to the lower electron-capture rate during the
initial deflagration. However, increasing $\rho_\mathrm{c}$ to
$5.5 \times 10^9$\,\gcc\ results in a similar \nife\ yield
  as in the base model with $\rho_\mathrm{c} = 2.9 \times 10^9$\,\gcc,
but the yield of heavier stable Ni isotopes more than doubles (see
Table~\ref{tab:all}).

For completeness we show the widely used 1D deflagration model W7 of
\cite{W7} as computed by \cite{Mori/etal:2018} with updated
electron-capture rates. In this model, the deflagration front is
artificially accelerated from 8 to 30\%\ of the local sound speed.
The initial propagation of the deflagration flame results in a similar
electron-capture rate compared to the most energetic 3D deflagration
models of \cite{Fink/etal:2014}, with a \nife\ yield of $\sim
0.07$\,\msun. However the gradual acceleration of the flame results in
a more complete burn of the outer layers and a larger \nifs\ yield
compared to standard deflagration models. \cite{Leung/Nomoto:2018}
find similar \nifs\ and stable Ni yields but they also consider W7
models at sub-solar metallicities (0.1 and
0.5\,Z$_\odot$). Interestingly, the impact on the stable Ni yield is
negligible ($\lesssim 4$\%; see Table~\ref{tab:all}).

\paragraph{\textbf{Delayed detonations:}}

In the 1D delayed-detonation models shown here (DDC series of
\citealt{Blondin/etal:2013}), the \nife\ yield is relatively constant
at $\sim 0.025$\,\msun\ regardless of the \nifs\ mass. Stable Ni
isotopes are almost exclusively synthesized in high-density burning
conditions during the early deflagration phase, with almost no stable
Ni produced during the subsequent detonation phase (where most of the
radioactive \nifs\ is synthesized).  The exception is model DDC0 which
has the largest \nifs\ yield (0.84\,\msun) for which an additional
$\sim 0.01$\,\msun\ of \nife\ is synthesized during the detonation
phase at expansion velocities
$5000\mathrm{-}8500$\,\kms\ (corresponding to mass coordinates $\sim
0.3\mathrm{-}0.65$\,\msun), where the peak temperatures reach
$5.5\mathrm{-}6.9\times 10^9$\,K at densities
$2.5\mathrm{-}5.8\times 10^7$\,\gcc.

The situation is somewhat different in 3D simulations where a
substantial amount of stable Ni is synthesized during the
detonation phase. The weaker the initial deflagration (i.e. the lower
the number of ignition kernels), the smaller the WD pre-expansion prior to
the deflagration-to-detonation transition and the higher the burning
density during the detonation phase. As a result, more stable IGEs (as
well as radioactive \nifs) are synthesized during the detonation. An
extreme example is model N1 (only one ignition kernel ignites 
the initial deflagration), which synthesizes more than 0.07\,\msun\ of
\nife\ and more than 1.1\,\msun\ of \nifs, but whose deflagration
counterpart (N1def) synthesizes less than 0.01\,\msun\ of \nife\ and
less than 0.1\,\msun\ of \nifs. Conversely, one can deduce from
comparing models N1600 and N1600def that almost all the stable
\nife\ and radioactive \nifs\ are synthesized during the deflagration
phase, owing to the high number of ignition kernels.

Also shown in Fig.~\ref{fig:mni58} is the impact of varying the
central density of the progenitor WD for model N100 (dashed line; see
also the 2D models of \citealt{Kobayashi/etal:2020}). As
for the deflagration model N100def, a lower central density of
$\rho_\mathrm{c} = 1.0 \times 10^9$\,\gcc\ results in a lower
\nife\ yield owing to the lower electron-capture rate during the
initial deflagration. However, whereas increasing $\rho_\mathrm{c}$
had a negligible impact on the production of stable \nife\ for the
deflagration model N100def, the \nife\ yield increases by $\sim
9$\%\ in the delayed-detonation model N100 owing to
pockets of high-density fuel burnt during the subsequent detonation
phase.

Finally, the impact of decreasing the metallicity of the
progenitor WD to one half, one tenth, and one hundredth solar is shown
for model N100 (dotted line). As expected, decreasing the metallicity
(and hence increasing \ye) favours the synthesis of radioactive
\nifs\ at the expense of stable \nife\ \citep[see
  e.g.][]{Timmes/etal:2003}.

\cite{Kobayashi/etal:2020} recomputed the 2D delayed-detonation models
of \cite{Leung/Nomoto:2018} by assuming a solar-scaled initial
composition as a proxy for the progenitor metallicity.  In
Fig.~\ref{fig:mni58}, we show their $Z=0.02$ models for three different
central densities corresponding to WD masses of 1.33, 1.37, and
1.38\,\msun\ (from low to high \nife\ yield; right half-filled circles
connected with a dashed line and labelled `zscl' in
Table~\ref{tab:all}). As noted in Sect.~\ref{sect:nse}, this results in
a much lower \nett\ mass fraction at a given metallicity compared to
what is expected from the conversion of CNO into \nett. This largely
explains the lower \nife\ yields compared to the models of
\cite{Seitenzahl/etal:2013}. \cite{Kobayashi/etal:2020} also present
the original models of \cite{Leung/Nomoto:2018} in which the
\nett\ mass fraction was set to the progenitor metallicity (labelled
`zne22' in Table~\ref{tab:all}). In Fig.~\ref{fig:mni58}, we show their
1.33\,\msun\ and 1.38\,\msun\ models at $Z=X(\nett)=0.02$. The
\nife\ yield is larger by up to a factor of three compared to the
corresponding solar-scaled initial composition models (connected via a
dotted line). We present models from \cite{Kobayashi/etal:2020} at
different metallicities in Table~\ref{tab:all}.

\paragraph{\textbf{Gravitationally confined detonation (GCD):}}
 
In this model, originally proposed by \cite{Plewa/etal:2004}, burning
is initiated as a weak central deflagration which drives a buoyant
bubble of hot ash that breaks out at the stellar surface, causing a
lateral acceleration and convergence of the flow of material at the
opposite end. Provided the density of the compressed material is high
enough, a detonation is triggered which incinerates the remainder of
the WD.

In Fig.~\ref{fig:mni58}, we show the 3D GCD model of
\cite{Seitenzahl/etal:2016} (half-filled pentagon), with a
\nife\ yield of $0.037$\,\msun\ for a \nifs\ yield of $0.742$\,\msun.
The weak initial deflagration results in little WD pre-expansion. In
this respect, it is similar to the delayed-detonation models of
\cite{Seitenzahl/etal:2013} with a low number of ignition spots, where
a significant amount of stable IGEs and \nifs\ are synthesized during
the detonation phase. However, the WD does expand during the flow
convergence phase, so less \nife\ is synthesized compared to
delayed-detonation models with similar \nifs\ yield.

\subsubsection{Sub-Chandrasekhar-mass models}

\paragraph{\textbf{Detonations:}}

For detonations of sub-\mch\ WDs the main parameter that determines
the final yields is the mass of the exploding WD. The propagation of
the detonation front is so fast compared to the WD expansion timescale
that the density at which material is burnt is close to the original
density profile of the progenitor WD. Lower-mass WDs have lower
densities at a given mass (or radial) coordinate, so the detonation
produces less electron-capture isotopes than for more massive WDs. For
the 1D sub-\mch\ models at solar metallicity shown here (SCH series of
\citealt{Blondin/etal:2017}; filled diamonds in Fig.~\ref{fig:mni58}),
only the highest-mass WDs ($M_\mathrm{WD} > 1.1$\,\msun) have a
\nife\ yield comparable to 1D delayed-detonation models (DDC series of
\citealt{Blondin/etal:2013}).
For WD masses below 1\,\msun, the \nife\ yield is significantly lower
($<0.025$\,\msun) yet not vanishingly small. Stable \nife\ is still
synthesized in detonations of $\lesssim 0.90$\,\msun\ WDs that result
in low-luminosity \sneia\ \citep[e.g.][]{Blondin/etal:2018}.

Varying the initial metallicity has the same effect as for the
\mch\ delayed-detonation models discussed above. In the
1\,\msun\ models of \cite{Shen/etal:2018}, increasing the metallicity
from solar to twice solar results in a factor of $\sim 2$ increase in
the \nife\ yield (from $7.05\times 10^{-3}$\,\msun\ to $1.66 \times
10^{-2}$\,\msun), whereas decreasing the metallicity from solar to
one-half solar results in a factor of $\sim 3$ decrease in the
\nife\ yield (from $7.05\times 10^{-3}$\,\msun\ to $2.48 \times
10^{-3}$\,\msun). Similar trends are observed for the slightly
super-solar ($\sim$1.6\,\zsun) 1.06\,\msun\ model of
\cite{Bravo/etal:2019} and in the extensive set of 1D sub-\mch\ models
published by \cite{Kushnir/etal:2020}\footnote{We only show a subset
  of the 470 models presented in this study to illustrate the
  metallicity dependence of the \nife\ yield in models with a similar
  setup (model IDs: 13, 49, 82, 113, 140, 157--161, 174, 210, 243,
  274, 301, and 318--322). Further models with varying \nett\ initial
  mass fraction and initial C/O ratio in the progenitor WD are
  reported in Table~\ref{tab:all}, as well as models in which weak
  reactions are included (labelled `CIWD\_NNNw'; the impact on the
  stable Ni yields is negligible).}.  We note that \nife\ is still
synthesized at zero metallicity in these models (with a yield $\sim
10^{-3}$\,\msun; see Table~\ref{tab:all}).

Several sub-\mch\ models at super-solar metallicities have higher
\nife\ (and total stable Ni) yields compared to some of the
delayed-detonation models shown here, such as the
3\,$\mathrm{Z}_\odot$ 1.06\,\msun\ model of
\cite{Sim/etal:2010}\footnote{The other models of \cite{Sim/etal:2010}
  are at zero metallicity, and, hence their \nife\ yield is less than
  0.002\,\msun.} and the 2\,$\mathrm{Z}_\odot$ 1.1\,\msun\ model of
\cite{Kushnir/etal:2020}, which yield $\sim 0.05$\,\msun\ and $\sim
0.04$\,\msun\ of \nife, respectively.

\paragraph{\textbf{Double detonations:}}
These models include a thin accreted helium layer which serves as a
trigger for detonating the underlying CO core. Since the
nucleosynthesis of \nife\ largely occurs in the CO core, its abundance
is expected to be similar to detonations of sub-\mch\ WDs for a given
WD mass. For instance, the 2D double-detonation model of
\cite{Townsley/etal:2019} from a 1\,\msun\ WD progenitor with a
0.021\,\msun\ He shell has very similar \nifs\ and \nife\ yields
compared to the 1\,\msun\ solar-metallicity model of
\cite{Shen/etal:2018}. The 3D models of \cite{Gronow/etal:2021}
display a quasi-linear trend of increasing \nife\ yield with
increasing \nifs\ mass (and hence progenitor WD mass), with a slight
offset to higher \nife\ yields compared to the 1D models of
\cite{Blondin/etal:2017}.  For clarity we do not show the
zero-metallicity models of \cite{Gronow/etal:2020} based on
1.05\,\msun\ progenitors as they produce a cluster of points around
$M(\nifs)\approx 0.6$\,\msun\ and $M(\nife)\approx 10^{-3}$\,\msun,
although we do include them in Table~\ref{tab:all}.  We do not show
results from the 2D double-detonation models of \cite{Fink/etal:2010}
as the corresponding abundance data is not available (R\"opke 2020,
private communication).

Owing to their prescription for the progenitor metallicity (see
Sect.~\ref{sect:nse}), the 2D double-detonation models of
\cite{Kobayashi/etal:2020} with solar-scaled initial composition for
$Z=0.02$ (left half-filled circles in Fig.~\ref{fig:mni58} and
labelled `zscl' in Table~\ref{tab:all}) have \nife\ yields of a few
$10^{-3}$\,\msun\ at most, comparable to zero-metallicity
sub-\mch\ models published by other groups
\citep[e.g.][]{Sim/etal:2010,Kushnir/etal:2020,Gronow/etal:2020}.  We
also show their 1\,\msun\ model at $Z=0.02$ in which the \nett\ mass
fraction was set to the initial metallicity (i.e. $X(\nett)=0.02$,
labelled `zne22' in Table~\ref{tab:all}). The \nife\ yield is one
order of magnitude larger compared to the corresponding solar-scaled
initial composition model (connected via a dotted line), and the total
stable Ni yield is larger by a factor of $\sim 3$.
  
\paragraph{\textbf{Detonations in ONe WDs:}}
In the 2D simulations carried out by \cite{Marquardt/etal:2015} the
progenitor ONe WDs are in the mass range
$1.18\mathrm{-}1.25$\,\msun\ with corresponding central densities
$1.0\mathrm{-}2.0 \times 10^8$\,\gcc, which results in the production
of copious amounts of \nifs\ ($>0.8$\,\msun). The initial composition includes
$^{20}$Ne but no \nett, hence, the \nife\ yield is low ($<5\times
10^{-3}$\,\msun), comparable to
other zero-metallicity models shown in Fig.~\ref{fig:mni58}.

\paragraph{\textbf{Violent WD mergers:}}
In the violent merger of two sub-\mch\ WDs, the nucleosynthesis of
IGEs occurs in similar conditions compared to detonations of single
sub-\mch\ WDs. The secondary (accreted) WD is almost entirely burned
in the process but at significantly lower densities, producing
intermediate-mass elements from incomplete silicon burning and oxygen
from carbon burning, while leaving some unburnt CO fuel.  Of the four
violent merger models with published nucleosynthesis data, solely the
model of \cite{Pakmor/etal:2012} corresponding to the violent merger
of two CO WDs of 0.9\,\msun\ and 1.1\,\msun\ has a significant
\nife\ yield ($\sim 0.028$\,\msun). The other three models have either
overly low metallicity ($0.9+0.9$\,\msun\ from the model of
\citealt{Pakmor/etal:2011} at zero metallicity;
$0.9+0.76$\,\msun\ model of \citealt{Kromer/etal:2016} at $Z=0.01$;
both yield a few times $10^{-5}$\,\msun\ of \nife) or reach too low
a peak density during the detonation ($\rho_\mathrm{peak}\lesssim
2\times 10^6$\,\gcc\ in the $0.9+0.76$\,\msun\ model of
\citealt{Kromer/etal:2013b}; the \nife\ yield is $\sim 0.002$\,\msun).

\paragraph{\textbf{WD-WD collisions:}}
Following the pioneering work of
\cite{Benz/etal:1989}, several groups have performed 3D simulations of
WD collisions with varying mass ratios and impact parameters
\citep{Raskin/etal:2009,Raskin/etal:2010,Rosswog/etal:2009,Loren-Aguilar/etal:2010,Hawley/etal:2012}.
However, all of these studies consider pure CO WDs (i.e. no \nett,
equivalent to zero metallicity in our framework), and none report
\nife\ yields due to their use of limited nuclear reaction networks
(the yield is expected to be low due to the zero metallicity, as in
the 2D simulations of \citealt{Papish/Perets:2016} who report
\nife\ yields $\lesssim 0.005$\,\msun\ for two of their models).

Here, we show the preliminary results of 2D simulations by Kushnir (2021,
private communication) consisting of equal-mass WD-WD collisions. From
low to high \nifs\ yield, the WD masses are: 0.5-0.5\,\msun,
0.6-0.6\,\msun, 0.7-0.7\,\msun, 0.8-0.8\,\msun, 0.9-0.9\,\msun, and
1.0-1.0\,\msun\ (not shown in Fig.~\ref{fig:mni58} for clarity,
although we do report its yields in Table~\ref{tab:all}). These
simulations extend the previous study of \cite{Kushnir/etal:2013} to
include a larger 69-isotope nuclear network and solar-metallicity WDs,
which results in sizeable \nife\ yields ($>10^{-2}$\,\msun\ for
collisions of WDs with masses of 0.7\,\msun\ and above; see
Table~\ref{tab:all}). The detonation conditions in WD collisions are
similar to those encountered in detonations of single sub-\mch\ WDs
(as is the case for the violent WD mergers discussed above), hence, the
stable \nife\ yields are similar at a given \nifs\ mass.

\subsubsection{Summary}

The \mch\ and sub-\mch\ models studied here clearly occupy distinct
regions of the $M(\nifs)$-$M(\nife)$ parameter space shown in
Fig.~\ref{fig:mni58}. At a given \nifs\ yield, and for the same
initial metallicity, sub-\mch\ models synthesize less \nife\ compared
to \mch\ models.

Typical \nife\ yields are 0--0.03\,\msun\ for sub-\mch\ models and
0.02--0.08\,\msun\ for \mch\ models (except for the weakest
\mch\ deflagration models N1def and N3def of \citealt{Fink/etal:2014}
which synthesize around 0.01\,\msun\ of \nife). This is modulated by
the progenitor metallicity and central density of the exploding WD. In
particular, reducing the central density by a factor of $\sim 3$
results in a $\sim 50$\% decrease in the \nife\ yield in the
delayed-detonation model N100 of \cite{Seitenzahl/etal:2013} and the
pure deflagration model N100def of \cite{Fink/etal:2014}.  The
synthesis of \nife\ does not necessarily require burning at the
highest central densities of \mch\ WD progenitors. The highest-mass
($M_\mathrm{WD} > 1$\,\msun) sub-\mch\ progenitors have \nife\ yields
comparable to some of the \mch\ models shown in Fig.~\ref{fig:mni58},
and sometimes even higher for super-solar metallicity progenitors.

The trend remains the same if we take into account the total stable
nickel yield as opposed to solely \nife. However, the double-detonation
models of \cite{Gronow/etal:2021} synthesize a significant fraction of
stable Ni in the form of $^{60}$Ni (20--30\%) and $^{62}$Ni
($\lesssim$10\%), and the double-detonation model of
\cite{Townsley/etal:2019} yields $\sim$45\% of stable Ni as $^{60}$Ni,
which causes these models to overlap with the 1D
\mch\ delayed-detonation models of \cite{Blondin/etal:2013}. Likewise,
the zero-metallicity double-detonation models of
\cite{Gronow/etal:2020} synthesize up to $\sim$90\% of their stable Ni
as $^{60}$Ni, resulting in an order of magnitude increase in their
stable Ni yields ($> 10^{-2}$\,\msun) compared to their \nife\ yields
($<2 \times 10^{-3}$\,\msun; see Table~\ref{tab:all}).

When considering the formation of [Ni\two] lines in late-time
\snia\ spectra ($\sim$200\,d past explosion in what follows), it is
the total stable Ni abundance at that time that matters. This
abundance is essentially set within the first day after the explosion,
as the sole decay chains with longer half-lives
($^{60}$Fe$\rightarrow^{60}$Co$\rightarrow ^{60}$Ni; see
Table~\ref{tab:decay2ni}) only contribute $\lesssim
10^{-4}$\,\msun\ of the total decayed stable Ni yield.  In the
following section, we explore whether the lower abundance of stable Ni
in sub-\mch\ models alone can explain the predicted lack of [Ni\two]
lines in their late-time spectra.


\section{Impact of stable Ni abundance and ionization on nebular [Ni\two]
  lines}\label{sect:ni2} 

\subsection{The absence of [Ni\two] lines from sub-\mch\ models}

\begin{figure}
\centering
\includegraphics{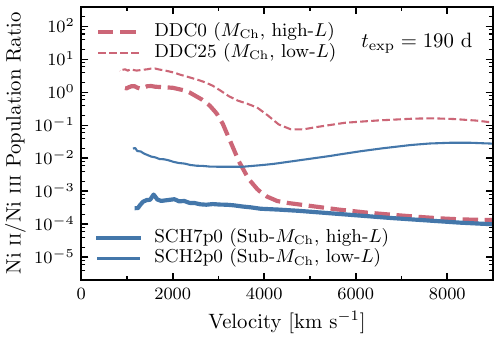}
\caption{\label{fig:ionni}
Ni\two/Ni\three\ population ratio at 190\,d past explosion for
  the high-luminosity models DDC0 (\mch; thick dashed line) and SCH7p0
(sub-\mch; thick solid line) as well as the low-luminosity models DDC25
(\mch; thin dashed line) and SCH2p0 (sub-\mch; thin solid
line). Regardless of the luminosity, Ni\three\ dominates in the
sub-\mch\ models all the way to the innermost ejecta ($\lesssim
3000$\,\kms), whereas Ni\two\ dominates there in the \mch\ models.
}
\end{figure}

In \cite{Blondin/etal:2018}, we concluded that the key parameter in
determining the presence of [Ni\two] lines in the late-time spectrum
of our low-luminosity \mch\ model DDC25 was larger abundance of stable Ni by a factor of $\sim17,$
compared to its sub-\mch\ counterpart
SCH2p0 ($2.9\times 10^{-2}$\,\msun\ in the \mch\ model cf. $1.7\times
10^{-3}$\,\msun\ in the sub-\mch\ model; see Table~\ref{tab:all}).
However, we also noted that the lower Ni ionization (i.e. higher
Ni\two/Ni\three\ ratio) in the inner ejecta of the \mch\ model further
enhanced their strength (Fig.~\ref{fig:ionni}, thin dashed line),
while the low Ni\two/Ni\three\ ratio in the sub-\mch\ model completely
suppressed both lines (Fig.~\ref{fig:ionni}, thin solid line; see also
\citealt{Wilk/etal:2018}).

\begin{figure}
\centering
\includegraphics{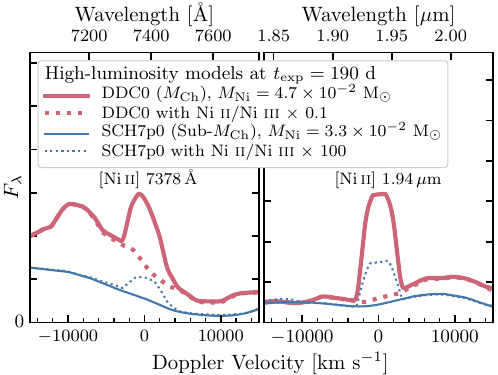}\vspace{.1cm}
\includegraphics{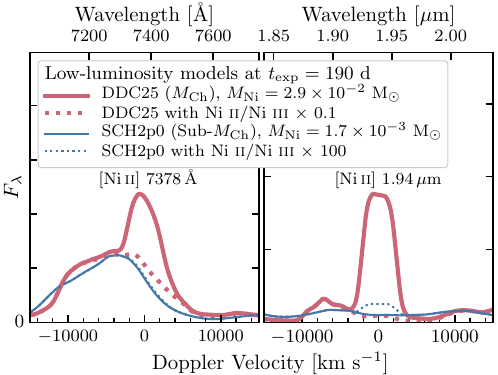}
\caption{\label{fig:ni2prof}
  Top panel:
  Optical (left) and near-infrared (right) [Ni\two] line
  profiles at 190\,d past explosion in the high-luminosity models DDC0
  (\mch; thick solid line) and SCH7p0 (sub-\mch; thin solid line).
  The dotted lines show the impact of artificially decreasing
  (increasing) the Ni\two/Ni\three\ ratio on the emergence of [Ni\two]
  lines in the \mch\ (sub-\mch) model.
  The near-infrared (NIR) line profiles were normalized to the same
  mean flux in the range $1.87\mathrm{-}1.88$\,$\mu$m; the optical
  profiles are not normalized.
  Bottom panel: Same as above for the low-luminosity models DDC25
  (\mch; thick solid line) and SCH2p0 (sub-\mch; thin solid line).
  We note the absence of an optical [Ni\two] 7378\,\AA\ line in the
  sub-\mch\ model with high Ni\two/Ni\three\ ratio see text
    for details).
}
\end{figure}

We  further explore the relative impact of abundance versus ionization on
the strength of [Ni\two] lines in late-time spectra by considering
\mch\ and sub-\mch\ models at the high-luminosity end, where the
differences in stable Ni yield are less pronounced (see
Fig.~\ref{fig:mni58}). For this, we use 
the \mch\ delayed-detonation model DDC0 and the sub-\mch\ detonation
model SCH7p0 (resulting from the detonation of a 1.15\,\msun\ WD
progenitor), both of which have a \nifs\ yield of $\sim
0.84$\,\msun. Unlike the aforementioned low-luminosity models, the
stable Ni yield is comparable in both models ($4.7\times
10^{-2}$\,\msun\ in the \mch\ model cf. $3.3\times
10^{-2}$\,\msun\ for the sub-\mch\ model).

Despite the similar stable Ni abundance, however, the ionization
profiles greatly differ and show the same behaviour as for the
low-luminosity models: Ni\two\ dominates in the inner ejecta of the
\mch\ model (Fig.~\ref{fig:ionni}, thick dashed line), whereas
Ni\three\ dominates in the sub-\mch\ model (Fig.~\ref{fig:ionni},
thick solid line). As a result, the \mch\ model displays prominent
[Ni\two] lines in its late-time spectrum, while the sub-\mch\ model
shows no such lines, as was the case for the low-luminosity models
(Fig.~\ref{fig:ni2prof}).

The higher Ni ionization in the sub-\mch\ models is a result of their
factor of 3-4 lower density in the inner $\sim$3000\,\kms,
which both lowers the Ni\three$\rightarrow${\sc ii} recombination rate
and increases the deposited decay energy per unit mass. The presence
of \nifs\ (which has all decayed to \cofs\ at 190\,d past explosion)
down to the central layers in these sub-\mch\ models causes the local
deposition of positron kinetic energy from \cofs\ decay to partly
compensate for the less efficient trapping of $\gamma$-rays: 40--50\%
of the deposited decay energy below 3000\,\kms\ is from positrons in
both models \citep[see also][]{Wilk/etal:2018}.

\subsection{Impact of the Ni\two/Ni\three\ ratio on [Ni\two] lines}

The presence of [Ni\two] lines thus appears to be mostly
related to an ionization effect. We illustrate this by
artificially increasing the Ni\two/Ni\three\ ratio (hence, decreasing
the ionization) of the sub-\mch\ models below 3000\,\kms, while keeping the
original stable Ni abundance and temperature profiles the same (see
Appendix~\ref{sect:ni23scl} for 
details on the numerical procedure).

The dominant formation mechanism for these lines is collisional
excitation, hence, their strength scales with the Ni\two\ population
density.  As a result, [Ni\two] lines do indeed emerge in both the
high-luminosity sub-\mch\ model (in which the stable Ni yield was
similar to the corresponding \mch\ model) and the low-luminosity
sub-\mch\ model (in which the stable Ni yield was a factor of $\sim
17$ lower than in the \mch\ model).  Despite being about six times
stronger than the NIR line\footnote{Since both transitions share the
same upper level (Table~\ref{tab:ni2}; see also
\citealt{Floers/etal:2020}), the ratio of the emergent flux in each
line only depends on the ratio of $\Delta E A_{ul}$, where $\Delta E$
is the transition energy and $A_{ul}$ is the Einstein coefficient for
spontaneous emission.}, the optical [Ni\two] 7378\,\AA\ line
only manages to produce a small excess flux in the low-luminosity
sub-\mch\ model SCH2p0, as it is swamped by the neighbouring [Ca\two]
7300\,\AA\ doublet. This does not occur in the high-luminosity
sub-\mch\ model SCH7p0 due to the lower Ca abundance in the inner
ejecta of this model ($X(\mathrm{Ca})<10^{-7}$ below
5000\,\kms\ cf. $5\mathrm{-}6\times10^{-2}$ in the low-luminosity
model SCH2p0).

Nonetheless, the emergent [Ni\two] lines in our sub-\mch\ models
remain comparatively weak compared to those in the \mch\ models, even
when the Ni\two/Ni\three\ ratio is increased by a factor of 100. This
is particularly true for the low-luminosity sub-\mch\ model SCH2p0,
which suggests that a Ni abundance of at least $10^{-2}$\,\msun\ is
needed to form strong [Ni\two] lines. This seemingly excludes
sub-\mch\ progenitors for low-luminosity \sneia\ presenting strong
[Ni\two] lines in their late-time spectra, at least in
solar-metallicity environments.

The question remains whether a physical mechanism exists to
 boost the Ni\two/Ni\three\ ratio in the inner ejecta of
sub-\mch\ models, which would cause [Ni\two] lines to emerge despite
the low Ni abundance. One possible mechanism is clumping: the higher
density in the clumps enhances the recombination rate, hence, reducing
the average ionization. Clumping is expected to result from
hydrodynamical instabilities during the initial deflagration phase of
\mch\ delayed-detonation models
\cite[e.g.][]{Golombek/Niemeyer:2005}. However, such instabilities are
not predicted in sub-\mch\ detonation models
\cite[e.g.][]{Garcia-Senz/etal:1999}. \cite{Mazzali/etal:2020}
has suggested that clumping could also develop at much later times
($\sim1.5$\,yr after explosion in their model for SN~2014J) through
the development of local magnetic fields, which could also occur in
sub-\mch\ ejecta. Clumping could also develop on an intermediate
timescale of days via the \nifs\ bubble effect
\citep[e.g.][]{WangChihYueh:2005}.  Regardless of its physical
origin, \cite{Wilk/etal:2020} found that clumping indeed lowers the
average ionization in the inner ejecta but not enough to produce a
Ni\two/Ni\three\ ratio favourable for the appearance of [Ni\two]
lines, even for a volume-filling factor $f=0.1$, which results in a
ten-fold increase of the density in the clumps.

Conversely, artificially decreasing the Ni\two/Ni\three\ ratio (hence,
increasing the Ni ionization) of the \mch\ models by a factor of 10
(while keeping the original stable Ni abundance the same) completely
suppresses both the optical and near-infrared [Ni\two] lines
(Fig.~\ref{fig:ni2prof}, thick dotted lines).  We
stress that this procedure is for illustrative purposes only since we
do not compute a proper radiative-transfer solution (in particular the
temperature profile is left unchanged, as are the population densities
of all other species).  In the following section, we investigate how
inward mixing of \nifs, as predicted in 3D delayed-detonation models,
could affect the appearance of [Ni\two] lines in \mch\ models.


\section{Impact of mixing on nebular [Ni\two] lines}\label{sect:mix}

\subsection{Macroscopic versus microscopic mixing}

Macroscopic mixing in \sneia\ occurs during the deflagration phase of
3D \mch\ delayed-detonation models, due to rising bubbles of buoyant
hot nuclear ash and downward mixing of nuclear fuel
\citep[e.g.][]{Seitenzahl/etal:2013}. In the innermost ejecta, stable
IGEs can be transported outwards while \nifs\ synthesized at higher
velocities is mixed inwards. As a result, there is no radial chemical
segregation between stable IGEs and \nifs-rich layers as in the 1D
\mch\ models studied here.  While it is not possible to simulate such
macroscopic mixing in 1D, where the composition is fixed at a given
radial (or velocity) coordinate, various numerical techniques have been
developed to approximate this and other multi-dimensional effects
\cite[see
  e.g.][]{Duffell:2016,Zhou:2017,Mabanta/Murphy:2018,Mabanta/etal:2019,Dessart/Hillier:2020}. 

Instead, a commonly used expedient in 1D consists in homogenizing the
composition in successive mass shells by applying a running boxcar
average \citep[e.g.][]{Woosley/etal:2007,Dessart/etal:2014b}. In this
approach, the mixing is both macroscopic (material is effectively
advected to larger and lower velocities) and microscopic (the composition
is completely homogenized within each mass shell at each step of the
running average). The method is convenient but results in non-physical
composition profiles that affect the spectral properties
\citep[e.g.][]{Dessart/Hillier:2020}.  We note that numerical
diffusion causes some level of microscopic mixing even in 3D
simulations.

Here, we simply wish to illustrate the impact of mixing on the strength
of [Ni\two] lines in late-time spectra of the \mch\ delayed-detonation
model DDC15 of \cite{Blondin/etal:2015}. For this, we adopt a fully
microscopic mixing approach by homogenizing the composition in the
inner ejecta below some cutoff velocity $\varv_\mathrm{mix}$. In what
follows, we refer to this as `uniform' mixing. The mass fraction of a
given species $i$ is set to its mass-weighted-average for $\varv \le
\varv_\mathrm{mix}$, and is left unchanged for $\varv >
\varv_\mathrm{orig} = \varv_\mathrm{mix} + \Delta
\varv_\mathrm{trans}$, where $\Delta\varv_\mathrm{trans} = \{500,
1000\}$\,\kms. To avoid strong compositional discontinuities at
$\varv_\mathrm{mix}$, we use a cosine function to smoothly transition
from the uniform to the unchanged composition over the interval
$[\varv_\mathrm{mix}, \varv_\mathrm{orig}]$. Formally, in each mass
shell with velocity coordinate $\varv$:

\begin{equation}
  X_i(\varv) = 
  \begin{cases}
    \dfrac
        {\sum\limits_{\varv' < \varv_\mathrm{mix}} X_i(\varv') \Delta M(\varv')}
        {\sum\limits_{\varv' < \varv_\mathrm{mix}} \Delta M(\varv')} &
        \text{for $\varv < \varv_\mathrm{mix}$}, \\
        X_i(\varv_\mathrm{mix}) + f_\mathrm{cos}\
        X_i(\varv_\mathrm{orig}) &
        \text{for $\varv_\mathrm{mix} \le \varv < \varv_\mathrm{orig}$}, \\
        X_i(\varv)\ \ \text{(unchanged)} &
        \text{for $\varv \ge \varv_\mathrm{orig}$}.
  \end{cases}
\end{equation}

\noindent
where

\begin{equation}
f_\mathrm{cos} = 
\dfrac{1}{2}
\left\{
1 - \cos
\left[
  \left(
  \dfrac{\varv - \varv_\mathrm{mix}}{\Delta \varv_\mathrm{trans}}
  \right)
  \pi
  \right]
\right\}.
\end{equation}

\noindent
This procedure conserves the total mass of each species as
the density profile is left unchanged.

\begin{figure}
\centering
\includegraphics{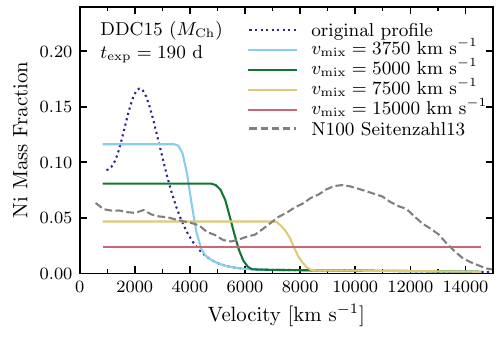}\vspace{.25cm}
\includegraphics{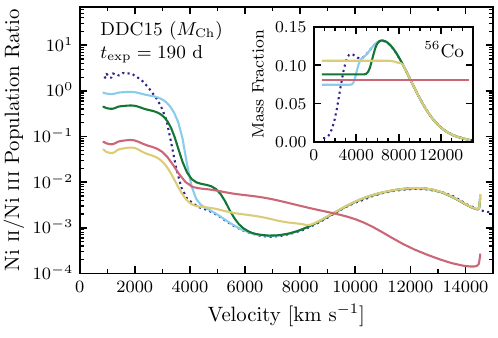}
\caption{\label{fig:mix}
Impact of uniformly mixing the
composition within a cutoff velocity $\varv_\mathrm{mix} = 3750, 5000, 7500$,
and $15000$\,\kms\ on the Ni abundance profile (top) and
Ni\two/Ni\three\ population ratio (bottom), illustrated using the
\mch\ delayed-detonation model DDC15 
of \cite{Blondin/etal:2015} at 190\,d past explosion. The
stable Ni mass for this model is $\sim 0.03$\,\msun\ (see
Table~\ref{tab:all}).
We show the angle-averaged Ni abundance profile of the 3D
delayed-detonation model N100 of \cite{Seitenzahl/etal:2013} for
comparison (grey dashed line, top panel). The inset in the lower
panel shows the \cofs\ abundance profiles, whose decay heating
by positrons and $\gamma$-rays largely
determines the ionization state at this time.
}
\end{figure}

The resulting Ni abundance profiles at 190\,d past explosion are shown
in Fig.~\ref{fig:mix} for values of $\varv_\mathrm{mix} =
3750,\ 5000,\ 7500$, and $15000$\,\kms\ (top panel). The
angle-averaged profile of the 3D delayed-detonation model N100 of
\cite{Seitenzahl/etal:2013} illustrates the advection of stable Ni to
larger velocities (grey dashed line), resulting in a stable Ni mass
fraction $\sim5\times10^{-2}$ below $\sim 4000$\,\kms, as in our
$\varv_\mathrm{mix}=7500$\,\kms\ model.

\begin{figure}
\centering
\includegraphics{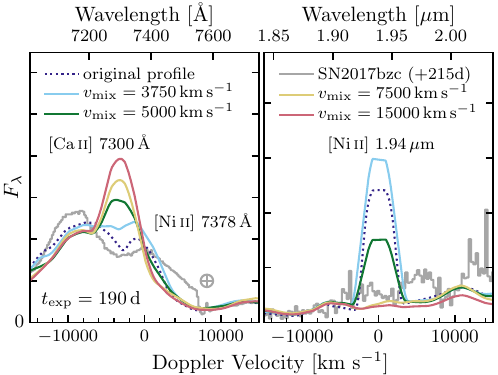}
\caption{\label{fig:ni2mix}
Impact of uniform mixing on the optical (left) and near-infrared
(right) [Ni\two] line profiles at 190\,d past explosion, using the
same models as in Fig.~\ref{fig:mix}. Also shown are observations of
\object{SN 2017bzc} at a slightly later phase (+215\,d past maximum) scaled to
match the mean flux of the original profile in the range
$7600\mathrm{-}8000$\,\AA\ and $1.83\mathrm{-}1.91$\,$\mu$m,
respectively (grey line). The feature marked with a `$\oplus$' at
+8000\,\kms\ in the optical spectrum is due to absorption by the
Earth's atmosphere (A-band). 
}
\end{figure}

\subsection{Impact of mixing on ionization and [Ni\two] lines}

The uniform mixing we apply not only affects the abundance profiles,
but the ionization as well (Fig.~\ref{fig:mix}, bottom panel). In the
inner 3000\,\kms, the Ni\two/Ni\three\ ratio systematically decreases
with increasing $\varv_\mathrm{mix}$. This increase in ionization
simply traces the increase in deposited energy from radioactive
decays, through inward mixing of \cofs\ (see inset). Unlike the
comparison between \mch\ and sub-\mch\ models in the previous section,
here the density profile is identical for all uniformly mixed models,
such that the \cofs\ radioactive decay heating (20-25\% of which is
due to local deposition of positron kinetic energy below $\sim
3000$\,\kms) predominantly determines the ionization state.

Variations in the amount of cooling through line emission further
affect the energy balance. This is best seen in the mixed model with
$\varv_\mathrm{mix} = 15000$\,\kms, where the Ni\two/Ni\three\ ratio
below $\sim 4000$\,\kms\ is comparable to the $\varv_\mathrm{mix} =
7500$\,\kms\ model despite the $\sim$20\%\ lower decay
heating\footnote{The mass fraction of \cofs\ is $\sim 0.08$ for the
  $\varv_\mathrm{mix} = 15000$\,\kms\ model cf. $\sim 0.11$ for the
  $\varv_\mathrm{mix} = 7500$\,\kms\ model in those layers.}. This is
due to the less efficient line cooling in the $\varv_\mathrm{mix} =
15000$\,\kms\ model (where [Ca\two] collisional cooling dominates due
to the larger Ca mass fraction in these layers) compared to the
$\varv_\mathrm{mix} = 7500$\,\kms\ model, in which cooling via
[Fe\two] and [Fe\three] transitions is more efficient.

The resulting [Ni\two] line profiles are shown in
Fig.~\ref{fig:ni2mix}. As expected, the [Ni\two] 1.94\,$\mu$m line is
only present in models where the Ni\two/Ni\three\ ratio fraction is
sufficiently high ($>10^{-1}$, that is, for $\varv_\mathrm{mix} = 3750$
and $5000$\,\kms, as well as in the original DDC15 model), and its
strength is modulated by the abundance of Ni in the line-formation
region. Thus, the $\varv_\mathrm{mix} = 3750$\,\kms\ model displays a
stronger [Ni\two] 1.94\,$\mu$m line compared to the original DDC15
model since the Ni mass fraction below $\sim 1500$\,\kms\ is
higher. The FWHM of the line is also slightly larger
($\sim$4500\,\kms\ cf. $\sim$4250\,\kms\ for the original profile) due
to the larger radial extension of the line-emission region. For
$\varv_\mathrm{mix} = 5000$\,\kms, the [Ni\two] 1.94\,$\mu$m line is
weaker than in the original unmixed model owing to the lower Ni mass
fraction below $\sim 3500$\,\kms. However, its FWHM is similar despite
the presence of stable Ni beyond 4000\,\kms, since the
Ni\two/Ni\three\ ratio drops below $10^{-1}$ in these layers.

This trend holds for the optical [Ni\two] 7378\,\AA\ line but is more
difficult to discern, as the [Ca\two] 7300\,\AA\ doublet progressively
emerges with increasing $\varv_\mathrm{mix}$. A weak [Ni\two]
7378\,\AA\ line is indeed present in the original DDC15 and in the
$\varv_\mathrm{mix} = 3750$\,\kms\ models, where the Ca mass fraction
is $< 10^{-2}$ below $\sim 3000$\,\kms. In the other models, inward
mixing of Ca results in a mass fraction of a few times $10^{-2}$ which
is sufficient to swamp the [Ni\two] 7378\,\AA\ line, as [Ca\two]
becomes a dominant coolant. The overabundance of Ca in the inner
ejecta illustrates a severe limitation of our 1D approach to mixing:
in the 3D delayed-detonation model N100 of
\cite{Seitenzahl/etal:2013}, the Ca mass fraction remains almost
systematically $\lesssim 10^{-5}$ below 5000\,\kms\ in all directions
(Seitenzahl 2021, private communication).

Furthermore, aside from low-luminosity 91bg-like events, the presence
of [Ca\two] 7300\,\AA\ in late-time spectra is not compatible with
observations of \sneia, as illustrated with SN~2017bzc in the left
panel of Fig.~\ref{fig:ni2mix} (grey line, \citealt{Floers/etal:2020};
see also \citealt{Maguire/etal:2018}). Our original (unmixed) DDC15
model does not predict significant [Ca\two] 7300\,\AA\ emission
(dotted line): The broad double-humped feature around 7300\,\AA\ is
dominated by [Ni\two] 7378\,\AA\ to the red and [Fe\two] 7155\,\AA\ to
the blue (as noted by \citealt{Floers/etal:2020}). However, our model
clearly overestimates the strength of [Ni\two] 1.94\,$\mu$m, and while
we can adjust the level of mixing to match its strength, this
inevitably results in an overestimation of [Ca\two] 7300\,\AA\ in the
optical.

Our results nonetheless suggest that inward mixing of \nifs\ can
completely wash out otherwise strong [Ni\two] lines in late-time
spectra of \mch\ models.  A more physical treatment of mixing could
result in pockets rich in stable nickel being physically isolated from
regions rich in \nifs\ (as in the 3D delayed-detonation models of
\citealt{Bravo/Garcia-Senz:2008}). This would suppress decay heating
of the stable Ni pockets by local positron kinetic energy deposition
from \cofs\ decay, and compensate in part for the increase in
ionization. Moreover, such stable Ni pockets could be moderately
compressed through the \nifs\ bubble effect
\citep[e.g.][]{WangChihYueh:2005,Dessart/etal:2021}, enhancing the Ni
recombination rate. Whatever the exact effect, mixing complicates the
use of [Ni\two] lines to constrain the stable Ni abundance, and the
absence of these lines cannot be unambiguously associated with a
sub-\mch\ explosion.


\section{Conclusions}\label{sect:ccl}

We studied the explosive nucleosynthesis of stable nickel and its
dominant isotope \nife\ in \sneia\ to test its use as a diagnostic of
the progenitor WD mass.  Among all reactions ending in \nife, we find
that the radiative proton-capture reaction
$^{57}\mathrm{Co}(\mathrm{p},\gamma)^{58}\mathrm{Ni}$ mostly
determines the final \nife\ abundance in both \mch\ and
sub-\mch\ models. Contrary to expectations, direct $\alpha$ captures
on \feff\ only contribute at the percent level to the net
nucleosynthetic flux to \nife, even in the $\alpha$-rich freeze-out
regime from nuclear statistical equilibrium.

At solar metallicity and for a given \nifs\ yield, sub-\mch\ models
synthesize less \nife\ ($\sim$0--0.03\,\msun) compared to \mch\ models
($\sim$0.02--0.08\,\msun), although this difference is reduced for WD
masses $\gtrsim 1$\,\msun\ or for super-solar metallicities.  The
trend remains the same when considering the total stable nickel yield
as opposed to only \nife, although some double-detonation models
synthesize 30--90\% of stable Ni in the form of heavier isotopes (in
particular $^{60}$Ni), causing an overlap with the 1D
\mch\ delayed-detonation models studied here.

The systematic absence of [Ni\two] lines in late-time spectra of
sub-\mch\ models is due to the higher Ni ionization in the inner
ejecta, where Ni\three\ dominates over Ni\two. This higher ionization
results from the lower density of the inner ejecta compared to
\mch\ models, which limits the Ni\three$\rightarrow${\sc ii}
recombination rate and increases the deposited decay energy per unit
mass. In 1D \mch\ models, the difference in ionization is exacerbated
by the under-abundance of \nifs\ in the inner ejecta, which results in
lower local kinetic energy deposition by positrons from \cofs\ decay
at late times.

Artificially reducing the Ni ionization of the sub-\mch\ models (while
maintaining the same Ni abundance) results in the emergence of
[Ni\two] lines, although these remain fairly weak in low-luminosity
models where the stable Ni yield is $< 10^{-2}$\,\msun, even when the
Ni\two/Ni\three\ ratio is increased by a factor of 100.  Any mechanism
that reduces the ionization state of the inner ejecta in
sub-\mch\ models could thus in principle lead to the formation of
[Ni\two] lines, thereby invalidating the use of this line as a
fool-proof discriminant between \mch\ and sub-\mch\ explosions. One
such mechanism is clumping, although a recent study by
\cite{Wilk/etal:2020} showed that the ionization level was not lowered
sufficiently to produce a favourable Ni\two/Ni\three\ ratio, at least
in their 1D implementation of clumping via volume-filling factors.

Likewise, an increase in the Ni ionization of the \mch\ models through
a tenfold reduction of the Ni\two/Ni\three\ ratio completely
suppresses both optical and near-infrared [Ni\two] lines, despite a
relatively large abundance of stable Ni ($3\mathrm{-}5\times
10^{-2}$\,\msun). This again demonstrates the importance of ionization
over abundance in determining the presence of [Ni\two] lines in
late-time spectra of \mch\ models.

Conversely, mixing can completely wash out otherwise strong [Ni\two]
lines in \mch\ models. Our investigation of this effect in 1D is
artificial, but nonetheless captures the main effect of the inward
microscopic mixing of \nifs\ and the resulting increase in decay
energy deposition and, hence, the Ni ionization state, in the inner
ejecta. At the same time, stable Ni is mixed outwards, reducing its
abundance in the [Ni\two] line-formation region.  A more elaborate
treatment of mixing could mitigate in part this increase in Ni
ionization.

In summary, the presence of [Ni\two] lines in late-time spectra of
\sneia\ is largely the result of a favourable Ni ionization state in
the inner ejecta and it is not guaranteed solely based on a large
abundance of stable nickel. This sensitivity to ionization complicates
the use of these lines alone as a diagnostic of the progenitor WD mass
(or simply differentiating between \mch\ and sub-\mch\ ejecta).  It is
possible that [Ni\two] lines in combination with other lines of
[Co\two/{\sc iii}] and [Fe\two/{\sc iii}] present in late-time spectra
could help constrain the Ni ionization state. In that case, a low Ni
ionization combined with an absence of [Ni\two] lines would point to a
very low abundance of stable nickel ($\lesssim 10^{-3}$\,\msun) and,
in turn, to a sub-\mch\ progenitor. Conversely, the presence of strong
[Ni\two] lines in a low-luminosity \snia\ would likely be the result
of a Chandrasekhar-mass explosion.


\begin{acknowledgements}
  The authors acknowledge useful discussions with Subo Dong, Chiaki
  Kobayashi, Doron Kushnir, Kate Maguire, Fritz R\"opke, Ivo
  Seitenzahl, Ken Shen, Kanji Mori, Dean Townsley, and members of the
  Garching SN group (in particular: Andreas Fl\"ors, Bruno Leibundgut,
  R\"udiger Pakmor, Jason Spyromilio, and Stefan Taubenberger).  SB
  thanks Inma Domínguez for performing the stellar-evolution
  calculation for model 5p0\_Z0p014, Chiaki Kobayashi and Shing Chi
  Leung for sending the tabulated yields from
  \cite{Kobayashi/etal:2020}, and Doron Kushnir for sending the nickel
  yields from his 2D equal-mass WD-WD collision models ahead of
  publication.  This work was supported by the `Programme National de
  Physique Stellaire' (PNPS) of CNRS/INSU co-funded by CEA and CNES.
  This research was supported by the Excellence Cluster ORIGINS which
  is funded by the Deutsche Forschungsgemeinschaft (DFG, German
  Research Foundation) under Germany's Excellence Strategy
  EXC-2094-390783311.  SB acknowledges support from the ESO Scientific 
  Visitor Programme in Garching.  EB's research is supported by MINECO
  grant PGC2018-095317-B-C21.  FXT's research is partially supported
  by the NSF under grant No. PHY-1430152 for the Physics Frontier
  Center Joint Institute for Nuclear Astrophysics Center for the
  Evolution of the Elements (JINA-CEE).  DJH thank NASA for partial
  support through theory grants NNX14AB41G and 80NSSC20K0524.  This
  research has made use of computing facilities operated by CeSAM data
  centre at LAM, Marseille, France.  This work made use of the
  Heidelberg Supernova Model Archive (HESMA),
  \url{https://hesma.h-its.org}.
\end{acknowledgements}


\bibliographystyle{aa} 
\bibliography{ms_ni2} 

\listofobjects

\begin{appendix}

\section{Basic model properties and nickel isotopic abundances}

Table~\ref{tab:all} gives basic properties and nickel isotopic
abundances of the models discussed in Sect.~\ref{sect:ni58yield} (see
also Fig.~\ref{fig:mni58}). Column headings are described hereafter:
\begin{itemize}

\item[(1)] model name
\item[(2)] dimension of the numerical simulation (1D, 2D, or 3D)

\item[(3)] total progenitor mass. We include the helium-shell mass for
  sub-\mch\ double-detonation models, whereas for the violent WD merger
  and WD-WD collision models, we give the total ejecta mass
  (equal to the combined mass of the two WDs)
\item[(4)] composition of the progenitor WD star prior to
  thermonuclear runaway. We only report the mass fractions of
  $^{12}$C, $^{16}$O, and $^{22}$Ne for the CO core (i.e. excluding
  the He shell for double-detonation models). The \nett\ mass fraction
  is almost always adjusted by hand to mimic a given metallicity, but
  the exact value assumed for a solar-metallicity WD ($X(\nett)\approx
  0.013$ according to Eq.~\ref{eq:xne22}) varies among different
  authors between 0.01 and 0.025. One exception is the \mch\
    delayed-detonation model 5p0\_Z0p014 published here for the first
    time. The \nett\ abundance (and C/O ratio) in this model
    results from a stellar-evolution calculation and takes into
    account the convective burning (or `simmering') phase prior to
    thermonuclear runaway as in \cite{Bravo/etal:2010}

\item[(5)] radioactive \nifs\ yield shortly after explosion
  ($t\approx 0$)

\item[(6)] decayed stable \nife\ yield at $t=1$\,yr past
  explosion. The \nife\ yield at $t=1$\,yr is essentially the same as
  at $t\approx 0$ (see Sect.~\ref{sect:ni58yield})
\item[(7)] total decayed stable Ni yield (including all stable isotopes:
  \nife, $^{60}$Ni, $^{61}$Ni, $^{62}$Ni, and $^{64}$Ni) at
  $t=1$\,yr past explosion. Since \nife\ is the only stable Ni isotope in
    the WD-WD collision models of Kushnir (2021, private communication), the
    total stable Ni yield is the same as the \nife\ yield
\item[(8)] main reference for the model
\end{itemize}

\onecolumn
{\footnotesize
\begin{longtable}[l]{l@{\hspace{1.8mm}}c@{\hspace{2.2mm}}c@{\hspace{2.2mm}}l@{\hspace{2.2mm}}c@{\hspace{2.2mm}}c@{\hspace{2.2mm}}c@{\hspace{2.2mm}}l}
\caption{Basic model properties and nickel isotopic abundances.}\label{tab:all} \\
\hline\hline
Model & \multicolumn{1}{c}{Dimension} & $M_{\mathrm{tot}}$ & \multicolumn{1}{c}{$X_\mathrm{init}$} & $M(^{56}\mathrm{Ni})_{t=0}$ & $M(^{58}\mathrm{Ni})_{t=1\ \mathrm{yr}}$ & $M(\mathrm{Ni}_{\mathrm{stable}})_{t=1\ \mathrm{yr}}$ & Reference \\
 &  & (M$_\odot$) &\multicolumn{1}{c}{\ $^{12}$C / $^{16}$O / $^{22}$Ne\ }& (M$_\odot$) & (M$_\odot$) & (M$_\odot$) &  \\
(1) & (2) & (3) & \multicolumn{1}{c}{(4)} & (5) & (6) & (7) & (8) \\
\hline
\endfirsthead
\caption{continued.} \\
\hline\hline
Model & \multicolumn{1}{c}{Dimension} & $M_{\mathrm{tot}}$ & \multicolumn{1}{c}{$X_\mathrm{init}$} & $M(^{56}\mathrm{Ni})_{t=0}$ & $M(^{58}\mathrm{Ni})_{t=1\ \mathrm{yr}}$ & $M(\mathrm{Ni}_{\mathrm{stable}})_{t=1\ \mathrm{yr}}$ & Reference \\
 &  & (M$_\odot$) &\multicolumn{1}{c}{\ $^{12}$C / $^{16}$O / $^{22}$Ne\ }& (M$_\odot$) & (M$_\odot$) & (M$_\odot$) &  \\
(1) & (2) & (3) & \multicolumn{1}{c}{(4)} & (5) & (6) & (7) & (8) \\
\hline
\endhead
\hline
\endfoot
\multicolumn{8}{c}{\it \mch\ Deflagrations}\\
N1def & 3D         & 1.40 & 0.475 / 0.500 / 2.5($-$2) & 0.063 & 7.41\,($-$3) & 7.78\,($-$3) & \cite{Fink/etal:2014} \\
N3def & 3D         & 1.40 & 0.475 / 0.500 / 2.5($-$2) & 0.084 & 1.31\,($-$2) & 1.40\,($-$2) & \cite{Fink/etal:2014} \\
N5def & 3D         & 1.40 & 0.475 / 0.500 / 2.5($-$2) & 0.174 & 2.39\,($-$2) & 2.52\,($-$2) & \cite{Fink/etal:2014} \\
N10def & 3D         & 1.40 & 0.475 / 0.500 / 2.5($-$2) & 0.197 & 3.00\,($-$2) & 3.17\,($-$2) & \cite{Fink/etal:2014} \\
N20def & 3D         & 1.40 & 0.475 / 0.500 / 2.5($-$2) & 0.266 & 4.20\,($-$2) & 4.48\,($-$2) & \cite{Fink/etal:2014} \\
N40def & 3D         & 1.40 & 0.475 / 0.500 / 2.5($-$2) & 0.335 & 5.54\,($-$2) & 5.89\,($-$2) & \cite{Fink/etal:2014} \\
N100def & 3D         & 1.40 & 0.475 / 0.500 / 2.5($-$2) & 0.360 & 6.07\,($-$2) & 6.48\,($-$2) & \cite{Fink/etal:2014} \\
N100Hdef & 3D         & 1.42 & 0.475 / 0.500 / 2.5($-$2) & 0.333 & 5.93\,($-$2) & 6.94\,($-$2) & \cite{Fink/etal:2014} \\
N100Ldef & 3D         & 1.36 & 0.475 / 0.500 / 2.5($-$2) & 0.330 & 3.23\,($-$2) & 3.29\,($-$2) & \cite{Fink/etal:2014} \\
N150def & 3D         & 1.40 & 0.475 / 0.500 / 2.5($-$2) & 0.385 & 6.53\,($-$2) & 6.97\,($-$2) & \cite{Fink/etal:2014} \\
N300Cdef & 3D         & 1.40 & 0.475 / 0.500 / 2.5($-$2) & 0.340 & 5.92\,($-$2) & 6.39\,($-$2) & \cite{Fink/etal:2014} \\
N200def & 3D         & 1.40 & 0.475 / 0.500 / 2.5($-$2) & 0.379 & 7.20\,($-$2) & 7.68\,($-$2) & \cite{Fink/etal:2014} \\
N1600def & 3D         & 1.40 & 0.475 / 0.500 / 2.5($-$2) & 0.347 & 7.39\,($-$2) & 7.98\,($-$2) & \cite{Fink/etal:2014} \\
N1600Cdef & 3D         & 1.40 & 0.475 / 0.500 / 2.5($-$2) & 0.320 & 6.12\,($-$2) & 6.80\,($-$2) & \cite{Fink/etal:2014} \\
W7 & 1D & 1.38 & 0.475 / 0.500 / 2.5($-$2) & 0.651 & 6.94 ($-$2) & 7.46 ($-$2) & \cite{Mori/etal:2018} \\
W7\_Z0.1 & 1D & 1.38 & 0.498 / 0.500 / 2.5($-$3) & 0.645 & 5.95 ($-$2) & 6.31 ($-$2) & \cite{Leung/Nomoto:2018} \\
W7\_Z0.5 & 1D & 1.38 & 0.488 / 0.500 / 1.3($-$2) & 0.651 & 5.98 ($-$2) & 6.34 ($-$2) & \cite{Leung/Nomoto:2018} \\
W7\_Zsun & 1D & 1.38 & 0.475 / 0.500 / 2.5($-$2) & 0.659 & 6.20 ($-$2) & 6.56 ($-$2) & \cite{Leung/Nomoto:2018} \\ 
\hline
\multicolumn{8}{c}{\it \mch\ Delayed Detonations}\\
N1 & 3D         & 1.40 & 0.475 / 0.500 / 2.5($-$2) & 1.110 & 7.26\,($-$2) & 7.53\,($-$2) & \cite{Seitenzahl/etal:2013} \\
N3 & 3D         & 1.40 & 0.475 / 0.500 / 2.5($-$2) & 1.040 & 6.78\,($-$2) & 7.13\,($-$2) & \cite{Seitenzahl/etal:2013} \\
N5 & 3D         & 1.40 & 0.475 / 0.500 / 2.5($-$2) & 0.974 & 6.98\,($-$2) & 7.39\,($-$2) & \cite{Seitenzahl/etal:2013} \\
N10 & 3D         & 1.40 & 0.475 / 0.500 / 2.5($-$2) & 0.939 & 7.13\,($-$2) & 7.58\,($-$2) & \cite{Seitenzahl/etal:2013} \\
N20 & 3D         & 1.40 & 0.475 / 0.500 / 2.5($-$2) & 0.778 & 6.63\,($-$2) & 7.15\,($-$2) & \cite{Seitenzahl/etal:2013} \\
N40 & 3D         & 1.40 & 0.475 / 0.500 / 2.5($-$2) & 0.655 & 6.89\,($-$2) & 7.39\,($-$2) & \cite{Seitenzahl/etal:2013} \\
N100 & 3D         & 1.40 & 0.475 / 0.500 / 2.5($-$2) & 0.604 & 6.90\,($-$2) & 7.40\,($-$2) & \cite{Seitenzahl/etal:2013} \\
N100H & 3D         & 1.42 & 0.475 / 0.500 / 2.5($-$2) & 0.694 & 7.54\,($-$2) & 8.72\,($-$2) & \cite{Seitenzahl/etal:2013} \\
N100L & 3D         & 1.36 & 0.475 / 0.500 / 2.5($-$2) & 0.532 & 3.81\,($-$2) & 3.91\,($-$2) & \cite{Seitenzahl/etal:2013} \\
N100\_Z0.01 & 3D         & 1.40 & 0.500 / 0.500 / 2.5($-$4) & 0.655 & 5.01\,($-$2) & 5.56\,($-$2) & \cite{Seitenzahl/etal:2013} \\
N100\_Z0.1 & 3D         & 1.40 & 0.498 / 0.500 / 2.5($-$3) & 0.649 & 5.09\,($-$2) & 5.65\,($-$2) & \cite{Seitenzahl/etal:2013} \\
N100\_Z0.5 & 3D         & 1.40 & 0.488 / 0.500 / 1.3($-$2) & 0.629 & 5.90\,($-$2) & 6.42\,($-$2) & \cite{Seitenzahl/etal:2013} \\
N150 & 3D         & 1.40 & 0.475 / 0.500 / 2.5($-$2) & 0.566 & 7.01\,($-$2) & 7.50\,($-$2) & \cite{Seitenzahl/etal:2013} \\
N200 & 3D         & 1.40 & 0.475 / 0.500 / 2.5($-$2) & 0.415 & 7.29\,($-$2) & 7.77\,($-$2) & \cite{Seitenzahl/etal:2013} \\
N300C & 3D         & 1.40 & 0.475 / 0.500 / 2.5($-$2) & 0.512 & 6.26\,($-$2) & 6.75\,($-$2) & \cite{Seitenzahl/etal:2013} \\
N1600 & 3D         & 1.40 & 0.475 / 0.500 / 2.5($-$2) & 0.364 & 7.48\,($-$2) & 8.07\,($-$2) & \cite{Seitenzahl/etal:2013} \\
N1600C & 3D         & 1.40 & 0.475 / 0.500 / 2.5($-$2) & 0.322 & 6.16\,($-$2) & 6.85\,($-$2) & \cite{Seitenzahl/etal:2013} \\
M1.33\_zscl\_z0     & 2D & 1.33 & 0.500 / 0.500 / 0.0 & 0.782 & 2.13\,($-$2) & 3.23\,($-$2) & \cite{Kobayashi/etal:2020} \\
M1.33\_zscl\_z0p002 & 2D & 1.33 & 0.499 / 0.499 / $\cdots$ $^{a}$ & 0.781 & 2.16\,($-$2) & 3.26\,($-$2) & \cite{Kobayashi/etal:2020} \\
M1.33\_zscl\_z0p01  & 2D & 1.33 & 0.495 / 0.495 / $\cdots$ $^{a}$ & 0.778 & 2.18\,($-$2) & 3.30\,($-$2) & \cite{Kobayashi/etal:2020} \\
M1.33\_zscl\_z0p02  & 2D & 1.33 & 0.490 / 0.490 / $\cdots$ $^{a}$ & 0.775 & 2.24\,($-$2) & 3.38\,($-$2) & \cite{Kobayashi/etal:2020} \\
M1.33\_zscl\_z0p04  & 2D & 1.33 & 0.480 / 0.480 / $\cdots$ $^{a}$ & 0.770 & 2.37\,($-$2) & 3.57\,($-$2) & \cite{Kobayashi/etal:2020} \\
M1.37\_zscl\_z0     & 2D & 1.37 & 0.500 / 0.500 / 0.0 & 0.680 & 4.29\,($-$2) & 5.58\,($-$2) & \cite{Kobayashi/etal:2020} \\
M1.37\_zscl\_z0p002 & 2D & 1.37 & 0.499 / 0.499 / $\cdots$ $^{a}$ & 0.678 & 4.32\,($-$2) & 5.61\,($-$2) & \cite{Kobayashi/etal:2020} \\
M1.37\_zscl\_z0p01  & 2D & 1.37 & 0.495 / 0.495 / $\cdots$ $^{a}$ & 0.675 & 4.36\,($-$2) & 5.66\,($-$2) & \cite{Kobayashi/etal:2020} \\
M1.37\_zscl\_z0p02  & 2D & 1.37 & 0.490 / 0.490 / $\cdots$ $^{a}$ & 0.673 & 4.42\,($-$2) & 5.74\,($-$2) & \cite{Kobayashi/etal:2020} \\
M1.37\_zscl\_z0p04  & 2D & 1.37 & 0.480 / 0.480 / $\cdots$ $^{a}$ & 0.669 & 4.51\,($-$2) & 5.88\,($-$2) & \cite{Kobayashi/etal:2020} \\
M1.38\_zscl\_z0     & 2D & 1.38 & 0.500 / 0.500 / 0.0 & 0.649 & 4.83\,($-$2) & 6.70\,($-$2) & \cite{Kobayashi/etal:2020} \\
M1.38\_zscl\_z0p002 & 2D & 1.38 & 0.499 / 0.499 / $\cdots$ $^{a}$ & 0.647 & 4.85\,($-$2) & 6.73\,($-$2) & \cite{Kobayashi/etal:2020} \\
M1.38\_zscl\_z0p01  & 2D & 1.38 & 0.495 / 0.495 / $\cdots$ $^{a}$ & 0.644 & 4.89\,($-$2) & 6.77\,($-$2) & \cite{Kobayashi/etal:2020} \\
M1.38\_zscl\_z0p02  & 2D & 1.38 & 0.490 / 0.490 / $\cdots$ $^{a}$ & 0.642 & 4.94\,($-$2) & 6.84\,($-$2) & \cite{Kobayashi/etal:2020} \\
M1.38\_zscl\_z0p04  & 2D & 1.38 & 0.480 / 0.480 / $\cdots$ $^{a}$ & 0.638 & 5.04\,($-$2) & 6.98\,($-$2) & \cite{Kobayashi/etal:2020} \\
M1.33\_zne22\_z0 & 2D & 1.33 & 0.500 / 0.500 / 0.0 & 0.845 & 2.13\,($-$2) & 3.23\,($-$2) & \cite{Kobayashi/etal:2020} \\
M1.33\_zne22\_z0p002 & 2D & 1.33 & 0.499 / 0.499 / 2.0\,($-$3) & 0.838 & 2.19\,($-$2) & 3.33\,($-$2) & \cite{Kobayashi/etal:2020} \\
M1.33\_zne22\_z0p01 & 2D & 1.33 & 0.495 / 0.495 / 1.0\,($-$2) & 0.750 & 3.15\,($-$2) & 4.34\,($-$2) & \cite{Kobayashi/etal:2020} \\
M1.33\_zne22\_z0p02 & 2D & 1.33 & 0.490 / 0.490 / 2.0\,($-$2) & 0.724 & 4.60\,($-$2) & 5.79\,($-$2) & \cite{Kobayashi/etal:2020} \\
M1.33\_zne22\_z0p04 & 2D & 1.33 & 0.480 / 0.480 / 4.0\,($-$2) & 0.678 & 7.49\,($-$2) & 8.70\,($-$2) & \cite{Kobayashi/etal:2020} \\
M1.37\_zne22\_z0 & 2D & 1.37 & 0.500 / 0.500 / 0.0 & 0.696 & 4.29\,($-$2) & 5.58\,($-$2) & \cite{Kobayashi/etal:2020} \\
M1.37\_zne22\_z0p002 & 2D & 1.37 & 0.499 / 0.499 / 2.0\,($-$3) & 0.689 & 4.37\,($-$2) & 5.69\,($-$2) & \cite{Kobayashi/etal:2020} \\
M1.37\_zne22\_z0p01 & 2D & 1.37 & 0.495 / 0.495 / 1.0\,($-$2) & 0.650 & 5.17\,($-$2) & 6.52\,($-$2) & \cite{Kobayashi/etal:2020} \\
M1.37\_zne22\_z0p02 & 2D & 1.37 & 0.490 / 0.490 / 2.0\,($-$2) & 0.627 & 6.36\,($-$2) & 7.71\,($-$2) & \cite{Kobayashi/etal:2020} \\
M1.37\_zne22\_z0p04 & 2D & 1.37 & 0.480 / 0.480 / 4.0\,($-$2) & 0.587 & 8.70\,($-$2) & 1.00\,($-$1) & \cite{Kobayashi/etal:2020} \\
M1.38\_zne22\_z0 & 2D & 1.38 & 0.500 / 0.500 / 0.0 & 0.675 & 4.83\,($-$2) & 6.70\,($-$2) & \cite{Kobayashi/etal:2020} \\
M1.38\_zne22\_z0p002 & 2D & 1.38 & 0.499 / 0.499 / 2.0\,($-$3) & 0.669 & 4.90\,($-$2) & 6.80\,($-$2) & \cite{Kobayashi/etal:2020} \\
M1.38\_zne22\_z0p01 & 2D & 1.38 & 0.495 / 0.495 / 1.0\,($-$2) & 0.620 & 5.65\,($-$2) & 7.58\,($-$2) & \cite{Kobayashi/etal:2020} \\
M1.38\_zne22\_z0p02 & 2D & 1.38 & 0.490 / 0.490 / 2.0\,($-$2) & 0.598 & 6.77\,($-$2) & 8.70\,($-$2) & \cite{Kobayashi/etal:2020} \\
M1.38\_zne22\_z0p04 & 2D & 1.38 & 0.480 / 0.480 / 4.0\,($-$2) & 0.560 & 8.98\,($-$2) & 1.09\,($-$1) & \cite{Kobayashi/etal:2020} \\
DDC0         & 1D         & 1.41 & 0.491 / 0.491 / 1.4($-$2) & 0.840 & 3.52\,($-$2) & 4.69\,($-$2) & \cite{Blondin/etal:2013} \\
DDC6         & 1D         & 1.41 & 0.491 / 0.491 / 1.4($-$2) & 0.709 & 2.63\,($-$2) & 3.04\,($-$2) & \cite{Blondin/etal:2013} \\
DDC10        & 1D         & 1.41 & 0.491 / 0.491 / 1.4($-$2) & 0.614 & 2.58\,($-$2) & 2.99\,($-$2) & \cite{Blondin/etal:2013} \\
DDC15        & 1D         & 1.41 & 0.491 / 0.491 / 1.4($-$2) & 0.507 & 2.56\,($-$2) & 2.97\,($-$2) & \cite{Blondin/etal:2013} \\
DDC17        & 1D         & 1.41 & 0.491 / 0.491 / 1.4($-$2) & 0.407 & 2.53\,($-$2) & 2.94\,($-$2) & \cite{Blondin/etal:2013} \\
DDC20        & 1D         & 1.41 & 0.491 / 0.491 / 1.4($-$2) & 0.297 & 2.51\,($-$2) & 2.92\,($-$2) & \cite{Blondin/etal:2013} \\
DDC22        & 1D         & 1.41 & 0.491 / 0.491 / 1.4($-$2) & 0.201 & 2.48\,($-$2) & 2.89\,($-$2) & \cite{Blondin/etal:2013} \\
DDC25        & 1D         & 1.41 & 0.491 / 0.491 / 1.4($-$2) & 0.125 & 2.43\,($-$2) & 2.85\,($-$2) & \cite{Blondin/etal:2013} \\
5p0\_Z0p014      & 1D         & 1.37 & 0.460 / 0.506 / 2.7($-$2) & 0.601 & 3.14\,($-$2) & 3.87\,($-$2) & This paper           \\
 \hline \multicolumn{8}{c}{\textit{\mch\ Gravitationally-Confined Detonations}} \\
 GCD200 & 3D & 1.40 & 0.475 / 0.500 / 2.5($-$2) & 0.742 & 3.74 ($-$2) & 3.95 ($-$2) & \cite{Seitenzahl/etal:2016} \\
\hline
\multicolumn{8}{c}{\it Sub-\mch\ Detonations}\\
SCH1p5       & 1D         & 0.88 & 0.491 / 0.491 / 1.4($-$2) & 0.080 & 1.21\,($-$3) & 1.48\,($-$3) & \cite{Blondin/etal:2017} \\
SCH2p0       & 1D         & 0.90 & 0.491 / 0.491 / 1.4($-$2) & 0.118 & 1.37\,($-$3) & 1.64\,($-$3) & \cite{Blondin/etal:2017} \\
SCH2p5       & 1D         & 0.93 & 0.491 / 0.491 / 1.4($-$2) & 0.172 & 1.55\,($-$3) & 1.79\,($-$3) & \cite{Blondin/etal:2017} \\
SCH3p0       & 1D         & 0.95 & 0.491 / 0.491 / 1.4($-$2) & 0.233 & 1.74\,($-$3) & 1.96\,($-$3) & \cite{Blondin/etal:2017} \\
SCH3p5       & 1D         & 0.98 & 0.491 / 0.491 / 1.4($-$2) & 0.306 & 1.97\,($-$3) & 2.17\,($-$3) & \cite{Blondin/etal:2017} \\
SCH4p0       & 1D         & 1.00 & 0.491 / 0.491 / 1.4($-$2) & 0.386 & 2.42\,($-$3) & 2.63\,($-$3) & \cite{Blondin/etal:2017} \\
SCH4p5       & 1D         & 1.03 & 0.491 / 0.491 / 1.4($-$2) & 0.470 & 5.66\,($-$3) & 7.33\,($-$3) & \cite{Blondin/etal:2017} \\
SCH5p0       & 1D         & 1.05 & 0.491 / 0.491 / 1.4($-$2) & 0.554 & 8.58\,($-$3) & 1.23\,($-$2) & \cite{Blondin/etal:2017} \\
SCH5p5       & 1D         & 1.08 & 0.491 / 0.491 / 1.4($-$2) & 0.637 & 1.17\,($-$2) & 1.66\,($-$2) & \cite{Blondin/etal:2017} \\
SCH6p0       & 1D         & 1.10 & 0.491 / 0.491 / 1.4($-$2) & 0.712 & 1.50\,($-$2) & 2.09\,($-$2) & \cite{Blondin/etal:2017} \\
SCH6p5       & 1D         & 1.13 & 0.491 / 0.491 / 1.4($-$2) & 0.778 & 2.09\,($-$2) & 2.74\,($-$2) & \cite{Blondin/etal:2017} \\
SCH7p0       & 1D         & 1.15 & 0.491 / 0.491 / 1.4($-$2) & 0.842 & 2.60\,($-$2) & 3.30\,($-$2) & \cite{Blondin/etal:2017} \\
1.00\_5050\_xsun & 1D         & 1.00 & 0.493 / 0.493 / 1.0($-$2) & 0.554 & 7.05\,($-$3) & 1.71\,($-$2) & \cite{Shen/etal:2018} \\
1.00\_5050\_z0p0 & 1D         & 1.00 &0.500 / 0.500 / 0.0 & 0.580 & 1.06\,($-$3) & 1.05\,($-$2) & \cite{Shen/etal:2018} \\
1.00\_5050\_z0p5 & 1D         & 1.00 & 0.497 / 0.497 / 5.0($-$3)  & 0.566 & 2.48\,($-$3) & 1.24\,($-$2) & \cite{Shen/etal:2018} \\
1.00\_5050\_z2p0 & 1D         & 1.00 & 0.487 / 0.487 / 2.0($-$2)  & 0.533 & 1.66\,($-$2) & 2.74\,($-$2) & \cite{Shen/etal:2018} \\
1p06\_Z2p25e-2         & 1D         & 1.06 & 0.476 / 0.498 / 2.2($-$2) & 0.657 & 2.41\,($-$2) & 3.16\,($-$2) & \cite{Bravo/etal:2019} \\
0p88\_Z2p25e-2         & 1D         & 0.88 & 0.476 / 0.498 / 2.2($-$2) & 0.167 & 2.86\,($-$3) & 3.20\,($-$3) & \cite{Bravo/etal:2019} \\
det\_0.81 & 1D         & 0.82 & 0.500 / 0.500 / 0.0 & 0.009 & 1.62\,($-$5) & 1.63\,($-$5) & \cite{Sim/etal:2010} \\
det\_0.88 & 1D         & 0.89 & 0.500 / 0.500 / 0.0 & 0.070 & 2.59\,($-$5) & 2.62\,($-$5) & \cite{Sim/etal:2010} \\
det\_0.97 & 1D         & 0.98 & 0.500 / 0.500 / 0.0 & 0.301 & 5.70\,($-$5) & 7.18\,($-$5) & \cite{Sim/etal:2010} \\
det\_1.06 & 1D         & 1.06 & 0.500 / 0.500 / 0.0 & 0.559 & 9.03\,($-$4) & 3.08\,($-$3) & \cite{Sim/etal:2010} \\
det\_1.06\_0.075Ne & 1D         & 1.06 & 0.425 / 0.500 / 7.5($-$2) & 0.434 & 5.22\,($-$2) & 5.23\,($-$2) & \cite{Sim/etal:2010} \\
det\_1.15 & 1D         & 1.15 & 0.500 / 0.500 / 0.0 & 0.809 & 1.75\,($-$3) & 5.99\,($-$3) & \cite{Sim/etal:2010} \\
CIWD\_13   & 1D & 0.80 & 0.500 / 0.500 / 0.0 & 0.055 & 2.22\,($-$5) & 2.35\,($-$5) & \cite{Kushnir/etal:2020} \\
CIWD\_324w & 1D & 0.80 & 0.500 / 0.500 / 0.0 & 0.053 & 2.61\,($-$5) & 2.70\,($-$5) & \cite{Kushnir/etal:2020} \\
CIWD\_157  & 1D & 0.80 & 0.496 / 0.496 / 7.5\,($-$3) & 0.036 & 4.87\,($-$4) & 4.87\,($-$4) & \cite{Kushnir/etal:2020} \\
CIWD\_416  & 1D & 0.80 & 0.491 / 0.491 / 1.4\,($-$2) & 0.026 & 9.42\,($-$4) & 1.24\,($-$3) & \cite{Kushnir/etal:2020} \\
CIWD\_415  & 1D & 0.80 & 0.492 / 0.492 / 1.4\,($-$2) & 0.026 & 9.38\,($-$4) & 1.18\,($-$3) & \cite{Kushnir/etal:2020} \\
CIWD\_174  & 1D & 0.80 & 0.492 / 0.492 / 1.5\,($-$2) & 0.027 & 9.39\,($-$4) & 9.41\,($-$4) & \cite{Kushnir/etal:2020} \\
CIWD\_364w & 1D & 0.80 & 0.492 / 0.492 / 1.5\,($-$2) & 0.026 & 9.32\,($-$4) & 9.34\,($-$4) & \cite{Kushnir/etal:2020} \\
CIWD\_438  & 1D & 0.80 & 0.693 / 0.292 / 1.5\,($-$2) & 0.049 & 1.06\,($-$3) & 1.07\,($-$3) & \cite{Kushnir/etal:2020} \\
CIWD\_433  & 1D & 0.80 & 0.292 / 0.693 / 1.5\,($-$2) & 0.015 & 5.71\,($-$4) & 5.71\,($-$4) & \cite{Kushnir/etal:2020} \\
CIWD\_318  & 1D & 0.80 & 0.485 / 0.485 / 3.0\,($-$2) & 0.018 & 1.75\,($-$3) & 1.76\,($-$3) & \cite{Kushnir/etal:2020} \\
CIWD\_49   & 1D & 0.85 & 0.500 / 0.500 / 0.0 & 0.144 & 2.46\,($-$5) & 2.63\,($-$5) & \cite{Kushnir/etal:2020} \\
CIWD\_332w & 1D & 0.85 & 0.500 / 0.500 / 0.0 & 0.138 & 2.76\,($-$5) & 2.86\,($-$5) & \cite{Kushnir/etal:2020} \\
CIWD\_158  & 1D & 0.85 & 0.496 / 0.496 / 7.5\,($-$3) & 0.129 & 6.55\,($-$4) & 6.56\,($-$4) & \cite{Kushnir/etal:2020} \\
CIWD\_420  & 1D & 0.85 & 0.491 / 0.491 / 1.4\,($-$2) & 0.110 & 1.33\,($-$3) & 1.54\,($-$3) & \cite{Kushnir/etal:2020} \\
CIWD\_419  & 1D & 0.85 & 0.492 / 0.492 / 1.4\,($-$2) & 0.113 & 1.33\,($-$3) & 1.50\,($-$3) & \cite{Kushnir/etal:2020} \\
CIWD\_210  & 1D & 0.85 & 0.492 / 0.492 / 1.5\,($-$2) & 0.121 & 1.34\,($-$3) & 1.34\,($-$3) & \cite{Kushnir/etal:2020} \\
CIWD\_372w & 1D & 0.85 & 0.492 / 0.492 / 1.5\,($-$2) & 0.114 & 1.33\,($-$3) & 1.33\,($-$3) & \cite{Kushnir/etal:2020} \\
CIWD\_439  & 1D & 0.85 & 0.693 / 0.292 / 1.5\,($-$2) & 0.169 & 1.48\,($-$3) & 1.48\,($-$3) & \cite{Kushnir/etal:2020} \\
CIWD\_434  & 1D & 0.85 & 0.292 / 0.693 / 1.5\,($-$2) & 0.048 & 8.65\,($-$4) & 8.66\,($-$4) & \cite{Kushnir/etal:2020} \\
CIWD\_319  & 1D & 0.85 & 0.485 / 0.485 / 3.0\,($-$2) & 0.114 & 2.68\,($-$3) & 2.69\,($-$3) & \cite{Kushnir/etal:2020} \\
CIWD\_82   & 1D & 0.90 & 0.500 / 0.500 / 0.0 & 0.284 & 3.35\,($-$5) & 3.56\,($-$5) & \cite{Kushnir/etal:2020} \\
CIWD\_340w & 1D & 0.90 & 0.500 / 0.500 / 0.0 & 0.276 & 2.96\,($-$5) & 3.12\,($-$5) & \cite{Kushnir/etal:2020} \\
CIWD\_159  & 1D & 0.90 & 0.496 / 0.496 / 7.5\,($-$3) & 0.267 & 8.50\,($-$4) & 8.50\,($-$4) & \cite{Kushnir/etal:2020} \\
CIWD\_424  & 1D & 0.90 & 0.491 / 0.491 / 1.4\,($-$2) & 0.249 & 1.78\,($-$3) & 1.93\,($-$3) & \cite{Kushnir/etal:2020} \\
CIWD\_423  & 1D & 0.90 & 0.492 / 0.492 / 1.4\,($-$2) & 0.251 & 1.77\,($-$3) & 1.89\,($-$3) & \cite{Kushnir/etal:2020} \\
CIWD\_435  & 1D & 0.90 & 0.292 / 0.693 / 1.5\,($-$2) & 0.193 & 1.53\,($-$3) & 1.53\,($-$3) & \cite{Kushnir/etal:2020} \\
CIWD\_440  & 1D & 0.90 & 0.693 / 0.292 / 1.5\,($-$2) & 0.306 & 2.00\,($-$3) & 2.00\,($-$3) & \cite{Kushnir/etal:2020} \\
CIWD\_243  & 1D & 0.90 & 0.492 / 0.492 / 1.5\,($-$2) & 0.259 & 1.80\,($-$3) & 1.80\,($-$3) & \cite{Kushnir/etal:2020} \\
CIWD\_380w & 1D & 0.90 & 0.492 / 0.492 / 1.5\,($-$2) & 0.251 & 1.78\,($-$3) & 1.79\,($-$3) & \cite{Kushnir/etal:2020} \\
CIWD\_320  & 1D & 0.90 & 0.485 / 0.485 / 3.0\,($-$2) & 0.250 & 3.78\,($-$3) & 3.79\,($-$3) & \cite{Kushnir/etal:2020} \\
CIWD\_113  & 1D & 1.00 & 0.500 / 0.500 / 0.0 & 0.570 & 3.11\,($-$3) & 6.56\,($-$3) & \cite{Kushnir/etal:2020} \\
CIWD\_348w & 1D & 1.00 & 0.500 / 0.500 / 0.0 & 0.564 & 1.40\,($-$3) & 2.27\,($-$3) & \cite{Kushnir/etal:2020} \\
CIWD\_160  & 1D & 1.00 & 0.496 / 0.496 / 7.5\,($-$3) & 0.552 & 3.20\,($-$3) & 7.11\,($-$3) & \cite{Kushnir/etal:2020} \\
CIWD\_428  & 1D & 1.00 & 0.491 / 0.491 / 1.4\,($-$2) & 0.532 & 7.09\,($-$3) & 8.41\,($-$3) & \cite{Kushnir/etal:2020} \\
CIWD\_427  & 1D & 1.00 & 0.492 / 0.492 / 1.4\,($-$2) & 0.533 & 7.10\,($-$3) & 8.47\,($-$3) & \cite{Kushnir/etal:2020} \\
CIWD\_441  & 1D & 1.00 & 0.693 / 0.292 / 1.5\,($-$2) & 0.573 & 9.98\,($-$3) & 1.60\,($-$2) & \cite{Kushnir/etal:2020} \\
CIWD\_274  & 1D & 1.00 & 0.492 / 0.492 / 1.5\,($-$2) & 0.539 & 8.47\,($-$3) & 1.25\,($-$2) & \cite{Kushnir/etal:2020} \\
CIWD\_388w & 1D & 1.00 & 0.492 / 0.492 / 1.5\,($-$2) & 0.533 & 7.25\,($-$3) & 8.55\,($-$3) & \cite{Kushnir/etal:2020} \\
CIWD\_436  & 1D & 1.00 & 0.292 / 0.693 / 1.5\,($-$2) & 0.498 & 6.68\,($-$3) & 8.64\,($-$3) & \cite{Kushnir/etal:2020} \\
CIWD\_321  & 1D & 1.00 & 0.485 / 0.485 / 3.0\,($-$2) & 0.519 & 2.01\,($-$2) & 2.46\,($-$2) & \cite{Kushnir/etal:2020} \\
CIWD\_140  & 1D & 1.10 & 0.500 / 0.500 / 0.0 & 0.827 & 8.16\,($-$3) & 1.78\,($-$2) & \cite{Kushnir/etal:2020} \\
CIWD\_356w & 1D & 1.10 & 0.500 / 0.500 / 0.0 & 0.825 & 6.88\,($-$3) & 1.62\,($-$2) & \cite{Kushnir/etal:2020} \\
CIWD\_161  & 1D & 1.10 & 0.496 / 0.496 / 7.5\,($-$3) & 0.809 & 6.10\,($-$3) & 1.66\,($-$2) & \cite{Kushnir/etal:2020} \\
CIWD\_432  & 1D & 1.10 & 0.491 / 0.491 / 1.4\,($-$2) & 0.790 & 1.69\,($-$2) & 2.67\,($-$2) & \cite{Kushnir/etal:2020} \\
CIWD\_431  & 1D & 1.10 & 0.492 / 0.492 / 1.4\,($-$2) & 0.791 & 1.68\,($-$2) & 2.66\,($-$2) & \cite{Kushnir/etal:2020} \\
CIWD\_301  & 1D & 1.10 & 0.492 / 0.492 / 1.5\,($-$2) & 0.792 & 1.71\,($-$2) & 2.75\,($-$2) & \cite{Kushnir/etal:2020} \\
CIWD\_396w & 1D & 1.10 & 0.492 / 0.492 / 1.5\,($-$2) & 0.791 & 1.73\,($-$2) & 2.71\,($-$2) & \cite{Kushnir/etal:2020} \\
CIWD\_442  & 1D & 1.10 & 0.693 / 0.292 / 1.5\,($-$2) & 0.812 & 1.81\,($-$2) & 3.02\,($-$2) & \cite{Kushnir/etal:2020} \\
CIWD\_437  & 1D & 1.10 & 0.292 / 0.693 / 1.5\,($-$2) & 0.769 & 1.61\,($-$2) & 2.46\,($-$2) & \cite{Kushnir/etal:2020} \\
CIWD\_322  & 1D & 1.10 & 0.485 / 0.485 / 3.0\,($-$2) & 0.762 & 4.07\,($-$2) & 5.10\,($-$2) & \cite{Kushnir/etal:2020} \\
ddet\_M1a & 3D         & 1.05 & 0.500 / 0.500 / 0.0 & 0.574 & 1.23\,($-$3) & 1.71\,($-$2) & \cite{Gronow/etal:2020} \\
ddet\_M2a & 3D         & 1.05 & 0.500 / 0.500 / 0.0 & 0.587 & 1.15\,($-$3) & 1.41\,($-$2) & \cite{Gronow/etal:2020} \\
ddet\_M2a\_13 & 3D     & 1.05 & 0.500 / 0.500 / 0.0 & 0.439 & 1.71\,($-$3) & 1.68\,($-$2) & \cite{Gronow/etal:2020} \\
ddet\_M2a\_21 & 3D     & 1.05 & 0.500 / 0.500 / 0.0 & 0.572 & 1.11\,($-$3) & 1.59\,($-$2) & \cite{Gronow/etal:2020} \\
ddet\_M2a\_36 & 3D     & 1.05 & 0.500 / 0.500 / 0.0 & 0.571 & 1.08\,($-$3) & 1.60\,($-$2) & \cite{Gronow/etal:2020} \\
ddet\_M2a\_79 & 3D     & 1.05 & 0.500 / 0.500 / 0.0 & 0.576 & 1.04\,($-$3) & 1.47\,($-$2) & \cite{Gronow/etal:2020} \\
ddet\_M2a\_i55 & 3D    & 1.05 & 0.500 / 0.500 / 0.0 & 0.601 & 1.18\,($-$3) & 1.49\,($-$2) & \cite{Gronow/etal:2020} \\
ddet\_M2b & 3D         & 1.05 & 0.500 / 0.500 / 0.0 & 0.588 & 1.15\,($-$3) & 1.40\,($-$2) & \cite{Gronow/etal:2020} \\
ddet\_M3a & 3D         & 0.91 & 0.500 / 0.500 / 0.0 & 0.337 & 1.15\,($-$3) & 8.18\,($-$3) & \cite{Gronow/etal:2020} \\
ddet\_M08\_03 & 3D         & 0.83 & 0.500 / 0.490 / 1.0($-$2) & 0.132 & 2.03\,($-$3) & 2.25\,($-$3) & \cite{Gronow/etal:2021} \\
ddet\_M08\_05 & 3D         & 0.86 & 0.500 / 0.490 / 1.0($-$2) & 0.201 & 3.27\,($-$3) & 3.64\,($-$3) & \cite{Gronow/etal:2021} \\
ddet\_M08\_10 & 3D         & 0.91 & 0.500 / 0.490 / 1.0($-$2) & 0.312 & 1.05\,($-$2) & 1.50\,($-$2) & \cite{Gronow/etal:2021} \\
ddet\_M08\_10\_r & 3D         & 0.91 & 0.500 / 0.490 / 1.0($-$2) & 0.327 & 8.06\,($-$3) & 1.29\,($-$2) & \cite{Gronow/etal:2021} \\
ddet\_M09\_03 & 3D         & 0.93 & 0.500 / 0.490 / 1.0($-$2) & 0.330 & 7.01\,($-$3) & 8.34\,($-$3) & \cite{Gronow/etal:2021} \\
ddet\_M09\_05 & 3D         & 0.95 & 0.500 / 0.490 / 1.0($-$2) & 0.386 & 1.05\,($-$2) & 1.46\,($-$2) & \cite{Gronow/etal:2021} \\
ddet\_M09\_10 & 3D         & 1.00 & 0.500 / 0.490 / 1.0($-$2) & 0.487 & 1.81\,($-$2) & 2.63\,($-$2) & \cite{Gronow/etal:2021} \\
ddet\_M09\_10\_r & 3D         & 1.00 & 0.500 / 0.490 / 1.0($-$2) & 0.503 & 1.62\,($-$2) & 2.73\,($-$2) & \cite{Gronow/etal:2021} \\
ddet\_M10\_02 & 3D         & 1.03 & 0.500 / 0.490 / 1.0($-$2) & 0.541 & 1.70\,($-$2) & 2.55\,($-$2) & \cite{Gronow/etal:2021} \\
ddet\_M10\_03 & 3D         & 1.06 & 0.500 / 0.490 / 1.0($-$2) & 0.591 & 2.05\,($-$2) & 3.15\,($-$2) & \cite{Gronow/etal:2021} \\
ddet\_M10\_05 & 3D         & 1.06 & 0.500 / 0.490 / 1.0($-$2) & 0.547 & 1.84\,($-$2) & 3.04\,($-$2) & \cite{Gronow/etal:2021} \\
ddet\_M10\_10 & 3D         & 1.11 & 0.500 / 0.490 / 1.0($-$2) & 0.762 & 2.61\,($-$2) & 4.26\,($-$2) & \cite{Gronow/etal:2021} \\
ddet\_M11\_05 & 3D         & 1.16 & 0.500 / 0.490 / 1.0($-$2) & 0.838 & 3.06\,($-$2) & 4.95\,($-$2) & \cite{Gronow/etal:2021} \\
M0.90\_zscl\_z0     & 2D & 0.95 & 0.500 / 0.500 / 0.0 & 0.029 & 6.86\,($-$5) & 6.49\,($-$4) & \cite{Kobayashi/etal:2020} \\
M0.90\_zscl\_z0p002 & 2D & 0.95 & 0.499 / 0.499 / $\cdots$ $^{a}$ & 0.028 & 2.11\,($-$4) & 8.34\,($-$4) & \cite{Kobayashi/etal:2020} \\
M0.90\_zscl\_z0p01  & 2D & 0.95 & 0.495 / 0.495 / $\cdots$ $^{a}$ & 0.027 & 4.27\,($-$4) & 1.12\,($-$3) & \cite{Kobayashi/etal:2020} \\
M0.90\_zscl\_z0p02  & 2D & 0.95 & 0.490 / 0.490 / $\cdots$ $^{a}$ & 0.025 & 5.98\,($-$4) & 1.45\,($-$3) & \cite{Kobayashi/etal:2020} \\
M0.90\_zscl\_z0p04  & 2D & 0.95 & 0.480 / 0.480 / $\cdots$ $^{a}$ & 0.023 & 8.89\,($-$4) & 2.03\,($-$3) & \cite{Kobayashi/etal:2020} \\
M1.00\_zscl\_z0     & 2D & 1.05 & 0.500 / 0.500 / 0.0 & 0.643 & 9.63\,($-$4) & 8.98\,($-$3) & \cite{Kobayashi/etal:2020} \\
M1.00\_zscl\_z0p002 & 2D & 1.05 & 0.499 / 0.499 / $\cdots$ $^{a}$ & 0.642 & 1.06\,($-$3) & 9.14\,($-$3) & \cite{Kobayashi/etal:2020} \\
M1.00\_zscl\_z0p01  & 2D & 1.05 & 0.495 / 0.495 / $\cdots$ $^{a}$ & 0.641 & 1.08\,($-$3) & 9.41\,($-$3) & \cite{Kobayashi/etal:2020} \\
M1.00\_zscl\_z0p02  & 2D & 1.05 & 0.490 / 0.490 / $\cdots$ $^{a}$ & 0.639 & 1.34\,($-$3) & 1.00\,($-$2) & \cite{Kobayashi/etal:2020} \\
M1.00\_zscl\_z0p04  & 2D & 1.05 & 0.480 / 0.480 / $\cdots$ $^{a}$ & 0.635 & 3.00\,($-$3) & 1.20\,($-$2) & \cite{Kobayashi/etal:2020} \\
M1.10\_zscl\_z0     & 2D & 1.15 & 0.500 / 0.500 / 0.0 & 0.861 & 1.13\,($-$3) & 1.13\,($-$2) & \cite{Kobayashi/etal:2020} \\
M1.10\_zscl\_z0p002 & 2D & 1.15 & 0.499 / 0.499 / $\cdots$ $^{a}$ & 0.860 & 1.47\,($-$3) & 1.17\,($-$2) & \cite{Kobayashi/etal:2020} \\
M1.10\_zscl\_z0p01  & 2D & 1.15 & 0.495 / 0.495 / $\cdots$ $^{a}$ & 0.859 & 1.60\,($-$3) & 1.21\,($-$2) & \cite{Kobayashi/etal:2020} \\
M1.10\_zscl\_z0p02  & 2D & 1.15 & 0.490 / 0.490 / $\cdots$ $^{a}$ & 0.857 & 2.39\,($-$3) & 1.32\,($-$2) & \cite{Kobayashi/etal:2020} \\
M1.10\_zscl\_z0p04  & 2D & 1.15 & 0.480 / 0.480 / $\cdots$ $^{a}$ & 0.853 & 4.60\,($-$3) & 1.57\,($-$2) & \cite{Kobayashi/etal:2020} \\
M1.20\_zscl\_z0     & 2D & 1.25 & 0.500 / 0.500 / 0.0 & 1.052 & 5.16\,($-$3) & 1.55\,($-$2) & \cite{Kobayashi/etal:2020} \\
M1.20\_zscl\_z0p002 & 2D & 1.25 & 0.499 / 0.499 / $\cdots$ $^{a}$ & 1.052 & 5.64\,($-$3) & 1.60\,($-$2) & \cite{Kobayashi/etal:2020} \\
M1.20\_zscl\_z0p01  & 2D & 1.25 & 0.495 / 0.495 / $\cdots$ $^{a}$ & 1.050 & 6.51\,($-$3) & 1.70\,($-$2) & \cite{Kobayashi/etal:2020} \\
M1.20\_zscl\_z0p02  & 2D & 1.25 & 0.490 / 0.490 / $\cdots$ $^{a}$ & 1.049 & 7.32\,($-$3) & 1.80\,($-$2) & \cite{Kobayashi/etal:2020} \\
M1.20\_zscl\_z0p04  & 2D & 1.25 & 0.480 / 0.480 / $\cdots$ $^{a}$ & 1.043 & 1.06\,($-$2) & 2.15\,($-$2) & \cite{Kobayashi/etal:2020} \\
M0.90\_zne22\_z0 & 2D & 0.95 & 0.500 / 0.500 / 0.0 & 0.026 & 6.86\,($-$5) & 6.49\,($-$4) & \cite{Kobayashi/etal:2020} \\
M0.90\_zne22\_z0p002 & 2D & 0.95 & 0.499 / 0.499 / 2.0\,($-$3) & 0.024 & 1.49\,($-$4) & 7.30\,($-$4) & \cite{Kobayashi/etal:2020} \\
M0.90\_zne22\_z0p01 & 2D & 0.95 & 0.495 / 0.495 / 1.0\,($-$2) & 0.018 & 4.52\,($-$4) & 1.04\,($-$3) & \cite{Kobayashi/etal:2020} \\
M0.90\_zne22\_z0p02 & 2D & 0.95 & 0.490 / 0.490 / 2.0\,($-$2) & 0.015 & 7.73\,($-$4) & 1.37\,($-$3) & \cite{Kobayashi/etal:2020} \\
M0.90\_zne22\_z0p04 & 2D & 0.95 & 0.480 / 0.480 / 4.0\,($-$2) & 0.013 & 1.08\,($-$3) & 1.72\,($-$3) & \cite{Kobayashi/etal:2020} \\
M1.00\_zne22\_z0 & 2D & 1.05 & 0.500 / 0.500 / 0.0 & 0.642 & 9.63\,($-$4) & 8.98\,($-$3) & \cite{Kobayashi/etal:2020} \\
M1.00\_zne22\_z0p002 & 2D & 1.05 & 0.499 / 0.499 / 2.0\,($-$3) & 0.638 & 1.19\,($-$3) & 9.78\,($-$3) & \cite{Kobayashi/etal:2020} \\
M1.00\_zne22\_z0p01 & 2D & 1.05 & 0.495 / 0.495 / 1.0\,($-$2) & 0.620 & 1.11\,($-$2) & 1.95\,($-$2) & \cite{Kobayashi/etal:2020} \\
M1.00\_zne22\_z0p02 & 2D & 1.05 & 0.490 / 0.490 / 2.0\,($-$2) & 0.600 & 2.50\,($-$2) & 3.31\,($-$2) & \cite{Kobayashi/etal:2020} \\
M1.00\_zne22\_z0p04 & 2D & 1.05 & 0.480 / 0.480 / 4.0\,($-$2) & 0.564 & 5.36\,($-$2) & 6.11\,($-$2) & \cite{Kobayashi/etal:2020} \\
M1.10\_zne22\_z0 & 2D & 1.15 & 0.500 / 0.500 / 0.0 & 0.861 & 1.13\,($-$3) & 1.13\,($-$2) & \cite{Kobayashi/etal:2020} \\
M1.10\_zne22\_z0p002 & 2D & 1.15 & 0.499 / 0.499 / 2.0\,($-$3) & 0.856 & 2.07\,($-$3) & 1.29\,($-$2) & \cite{Kobayashi/etal:2020} \\
M1.10\_zne22\_z0p01 & 2D & 1.15 & 0.495 / 0.495 / 1.0\,($-$2) & 0.835 & 1.62\,($-$2) & 2.66\,($-$2) & \cite{Kobayashi/etal:2020} \\
M1.10\_zne22\_z0p02 & 2D & 1.15 & 0.490 / 0.490 / 2.0\,($-$2) & 0.817 & 3.61\,($-$2) & 4.58\,($-$2) & \cite{Kobayashi/etal:2020} \\
M1.10\_zne22\_z0p04 & 2D & 1.15 & 0.480 / 0.480 / 4.0\,($-$2) & 0.751 & 7.44\,($-$2) & 8.32\,($-$2) & \cite{Kobayashi/etal:2020} \\
M1.20\_zne22\_z0 & 2D & 1.25 & 0.500 / 0.500 / 0.0 & 1.053 & 5.16\,($-$3) & 1.55\,($-$2) & \cite{Kobayashi/etal:2020} \\
M1.20\_zne22\_z0p002 & 2D & 1.25 & 0.499 / 0.499 / 2.0\,($-$3) & 1.048 & 7.08\,($-$3) & 1.78\,($-$2) & \cite{Kobayashi/etal:2020} \\
M1.20\_zne22\_z0p01 & 2D & 1.25 & 0.495 / 0.495 / 1.0\,($-$2) & 1.025 & 2.36\,($-$2) & 3.37\,($-$2) & \cite{Kobayashi/etal:2020} \\
M1.20\_zne22\_z0p02 & 2D & 1.25 & 0.490 / 0.490 / 2.0\,($-$2) & 0.995 & 4.53\,($-$2) & 5.45\,($-$2) & \cite{Kobayashi/etal:2020} \\
M1.20\_zne22\_z0p04 & 2D & 1.25 & 0.480 / 0.480 / 4.0\,($-$2) & 0.937 & 8.44\,($-$2) & 9.26\,($-$2) & \cite{Kobayashi/etal:2020} \\
ddet\_sm14\_d2e5 & 2D & 1.02 & 0.400 / 0.580 / 2.0($-$2) & 0.598 & 1.14 ($-$2) & 2.55 ($-$2) & \cite{Townsley/etal:2019} \\
det\_ONe10e7 & 2D         & 1.18 & 0.030 / 0.600 / 0.0 $^{b}$ & 0.832 & 1.89\,($-$3) & 3.49\,($-$3) & \cite{Marquardt/etal:2015} \\
det\_ONe13e7 & 2D         & 1.21 & 0.030 / 0.600 / 0.0 $^{b}$ & 0.941 & 2.81\,($-$3) & 2.81\,($-$3) & \cite{Marquardt/etal:2015} \\
det\_ONe15e7 & 2D         & 1.23 & 0.030 / 0.600 / 0.0 $^{b}$ & 0.957 & 3.26\,($-$3) & 5.12\,($-$3) & \cite{Marquardt/etal:2015} \\
det\_ONe17e7 & 2D         & 1.24 & 0.030 / 0.600 / 0.0 $^{b}$ & 0.990 & 3.74\,($-$3) & 5.64\,($-$3) & \cite{Marquardt/etal:2015} \\
det\_ONe20e7 & 2D         & 1.25 & 0.030 / 0.600 / 0.0 $^{b}$ & 1.030 & 4.43\,($-$3) & 6.36\,($-$3) & \cite{Marquardt/etal:2015} \\
\hline
\multicolumn{8}{c}{\it Violent WD Mergers}\\
09+09 & 3D         & 1.73 &0.500 / 0.500 / 0.0 & 0.124 & 2.90\,($-$5) & 2.96\,($-$5) & \cite{Pakmor/etal:2010} \\
11+09 & 3D         & 1.94 & 0.475 / 0.500 / 2.5($-$2) & 0.614 & 2.84\,($-$2) & 3.03\,($-$2) & \cite{Pakmor/etal:2012} \\
09+076\_Z1 & 3D         & 1.61 & 0.482 / 0.500 / 1.3($-$2) & 0.183 & 1.60\,($-$3) & 2.18\,($-$3) & \cite{Kromer/etal:2013b} \\
09+076\_Z0.01 & 3D         & 1.61 & 0.500 / 0.500 / 1.3($-$4) & 0.197 & 6.63\,($-$5) & 8.53\,($-$5) & \cite{Kromer/etal:2016} \\
 \hline \multicolumn{8}{c}{\textit{WD-WD Collisions}} \\
 0.5-0.5 & 2D & 1.00 & 0.493 / 0.493 / 1.5($-2$) & 0.171 & 2.60 ($-$3) & 2.60 ($-$3) & Kushnir (2021, priv. comm.) \\
 0.6-0.6 & 2D & 1.20 & 0.493 / 0.493 / 1.5($-2$)   & 0.378 & 6.20 ($-$3) & 6.20 ($-$3) & Kushnir (2021, priv. comm.) \\
 0.7-0.7 & 2D & 1.40 & 0.493 / 0.493 / 1.5($-2$)   & 0.620 & 1.16 ($-$2) & 1.16 ($-$2) & Kushnir (2021, priv. comm.) \\
 0.8-0.8 & 2D & 1.60 & 0.493 / 0.493 / 1.5($-2$)   & 0.723 & 1.46 ($-$2) & 1.46 ($-$2) & Kushnir (2021, priv. comm.) \\
 0.9-0.9 & 2D & 1.80 & 0.493 / 0.493 / 1.5($-2$)   & 0.779 & 1.21 ($-$2) & 1.21 ($-$2) & Kushnir (2021, priv. comm.) \\
 1.0-1.0 & 2D & 2.00 & 0.493 / 0.493 / 1.5($-2$)   & 1.206 & 3.68 ($-$2) & 3.68 ($-$2) & Kushnir (2021, priv. comm.) \\
\hline
\end{longtable}
 \flushleft \begin{footnotesize}
 \textbf{Note:} Numbers in parentheses correspond to powers of ten.\\
 $^{a}$ We were not able to confirm the exact \nett\ abundance in the solar-scaled composition models of \cite{Kobayashi/etal:2020}.\\
 $^{b}$ The oxygen-neon WDs considered in the \cite{Marquardt/etal:2015} study have the following initial composition: $X(^{12}\mathrm{C})=0.03$, $X(^{16}\mathrm{O})=0.6$, and  $X(^{20}\mathrm{Ne})=0.37$. Since $X(^{22}\mathrm{Ne})=0$, these models are considered to be at zero metallicity. \end{footnotesize}
}
\twocolumn

\section{Atomic data for [Ni\two] transitions}

In Table~\ref{tab:ni2} we give the atomic data for the optical
7378\,\AA\ and NIR 1.94\,$\mu$m [Ni\two] transitions used in our
\cmfgen\ calculations. 

\begin{table*}
\footnotesize
\caption{Forbidden [Ni\two] transitions used in our \cmfgen\ simulations. The
  data were obtained from \cite{Quinet/LeDourneuf:1996}.}\label{tab:ni2}
\begin{tabular}{rcccccc}
\hline
\multicolumn{1}{c}{$\lambda_\mathrm{air}$} & \multicolumn{2}{c}{Lower Level} & \multicolumn{2}{c}{Upper Level} & Oscillator strength $f$ & Einstein coefficient $A_{ul}$ \\
\multicolumn{1}{c}{(\AA)}                & Configuration & Index $l$        & Configuration & Index $u$       &                         & (s$^{-1}$)                   \\
\hline
 7377.829  & 3d$^9$ $^2$D$_e$[5/2]        & 1  & 3d$^8$(3F)4s $^2$F$_e$[7/2]  & 7  & $2.1719 \times 10^{-9}$  & $1.9950 \times 10^{-1}$ \\
19387.744  & 3d$^8$(3F)4s $^4$F$_e$[9/2]  & 3  & 3d$^8$(3F)4s $^2$F$_e$[7/2]  & 7  & $4.1701 \times 10^{-9}$  & $9.2450 \times 10^{-2}$ \\
\hline
\end{tabular}
\end{table*}

\section{Modifying the Ni\two/Ni\three\ ratio in \cmfgen}\label{sect:ni23scl}

In \cmfgen, the population density $n_l$ (in cm$^{-3}$) of any given
state (level) $l$ is determined via a solution to the time-dependent
statistical equilibrium equations \citep{Hillier/Dessart:2012}. The
population density of the ionization state $i+$ ($i=0$ for neutral,
$i=1$ for once-ionized etc.) for some species is then simply the sum
of $n_l$ over all $N$ levels of that ionization state:

\begin{equation}
n^{i+} = \sum_{l=1}^{N} {n_l}^{i+}.
\end{equation}

\noindent
We then define the Ni\two/Ni\three\ ratio as:

\begin{equation}\label{eqn:i23}
I_{23} = \frac{n^{+}}{n^{2+}}.
\end{equation}

In Sect.~\ref{sect:ni2} we artificially modify the
Ni\two/Ni\three\ ratio to test the impact on the resulting [Ni\two]
lines in late-time \snia\ spectra. At any given depth in the ejecta
(denoted by index $j$, which corresponds to a given radius or velocity
coordinate in our 1D spatial grid), we scale all the Ni\two\ and
Ni\three\ level population densities by a factor of $a_{1,j}$ and
$a_{2,j}$, respectively, to obtain new population densities:

\begin{equation}\label{eqn:n+}
\widetilde{n_j}^{+} = \sum_{l=1}^{N} a_{1,j}\ {n_{l,j}}^{+} = a_{1,j}\ {n_j}^{+}
\end{equation}

\noindent
and

\begin{equation}\label{eqn:n2+}
\widetilde{n_j}^{2+} = \sum_{l=1}^{N} a_{2,j}\ {n_{l,j}}^{2+} = a_{2,j}\ {n_j}^{2+}
\end{equation}

\noindent
in order to achieve a new Ni\two/Ni\three\ ratio
  $\widetilde{I_{23}}$ related to the original ratio
  $I_{23}$ by some pre-determined factor:

\begin{equation}\label{eqn:r23}
  \mathcal{R}_{23} = \frac{\widetilde{I_{23}}}{I_{23}}
  = \frac{\widetilde{n_j}^{+}}{\widetilde{n_j}^{2+}}\frac{{n_j}^{2+}}{{n_j}^{+}}
  = \frac{a_{1,j}}{a_{2,j}}.
\end{equation}

\begin{figure}
\centering
\includegraphics{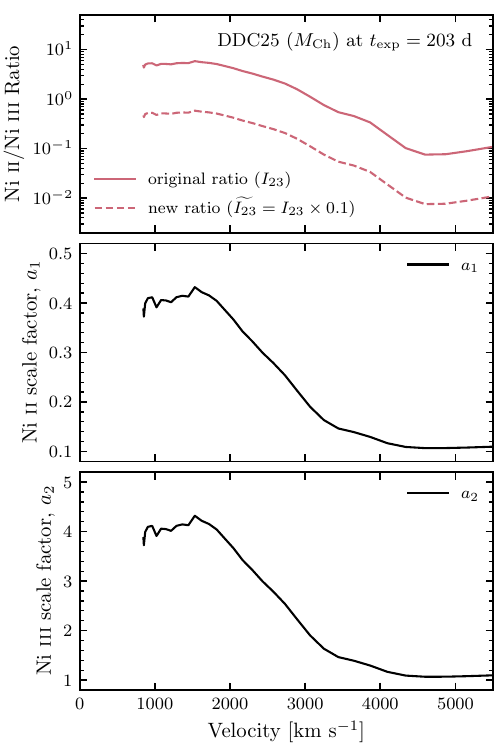}
\caption{\label{fig:popnick}
Illustration of the procedure used to modify the
Ni\two/Ni\three\ ratio in \cmfgen. In this example we wish to scale
the Ni\two/Ni\three\ ratio of the \mch\ model DDC25 at 203\,d past
explosion by $\mathcal{R}_{23}=0.1$. The upper panel shows the
original and scaled Ni\two/Ni\three\ ratios (Eq.~\ref{eqn:i23}). The
middle and bottom panels show the scaling coefficients $(a_1,a_2)$
applied to the population densities of Ni\two\ and Ni\three,
respectively (Eqs.~\ref{eqn:a1} and \ref{eqn:a2}). We note that
$a_1/a_2=\mathcal{R}_{23}=0.1$ at all depths by definition
(Eq.~\ref{eqn:r23}). When the original Ni\two/Ni\three\ ratio becomes
small enough ($\lesssim 10^{-1}$ beyond $\sim 4000$\,\kms),
$a_{1}\approx 0.1$ $(=\mathcal{R}_{23})$ and $a_{2}\approx 1$, as
expected (see text for details).
}
\end{figure}

Since we wish to preserve the total species population density at each
depth, we further require that:

\begin{equation}\label{eqn:popcons}
\widetilde{n_j}^{+} + \widetilde{n_j}^{2+} = {n_j}^{+} + {n_j}^{2+},
\end{equation}

\noindent
from which we derive an equation for the scale factor
for the Ni\three\ population density:

\begin{equation}\label{eqn:a2}
a_{2,j} = \frac{{n_j}^{+} (1-a_{1,j}) + {n_j}^{2+}}{{n_j}^{2+}},
\end{equation}

\noindent
which we then plug into Eq.~\ref{eqn:r23} to derive the scale
factor for the Ni\two\ population density:

\begin{equation}\label{eqn:a1}
  a_{1,j} = \frac{({n_j}^{+} + {n_j}^{2+})\ \mathcal{R}_{23}}
  {{n_j}^{+}\ \mathcal{R}_{23} + {n_j}^{2+}},
\end{equation}

\noindent
from which we trivially compute $a_{2,j} = a_{1,j}
\ \mathcal{R}_{23}$ using Eq.~\ref{eqn:r23} (or Eq.~\ref{eqn:a2}).
When ${n_j}^{+} \ll {n_j}^{2+}$ and $\mathcal{R}_{23} < 1$,
$a_{1,j}\approx \mathcal{R}_{23}$ and $a_{2,j}\approx 1$, as seen in
Fig.~\ref{fig:popnick}.

The scaled Ni\two\ and Ni\three\ population densities are used as an
input for an observer-frame calculation of the spectrum.

\end{appendix}

\end{document}